\documentclass[manuscript,screen]{acmart}

\AtBeginDocument{%
  }

\usepackage{longtable}
\usepackage{soul}
\usepackage{adjustbox}
\usepackage{graphicx}
\usepackage{multirow}
\usepackage{array} 
\usepackage{booktabs} 
\usepackage{makecell}
\usepackage{paralist}
\usepackage{amsmath}
\usepackage{mathtools}
\usepackage{xspace}
\usepackage[toc, acronym]{glossaries}
\newacronym{ML}{ML}{machine learning}
\newacronym{AI}{AI}{artificial intelligence}
\newacronym{DQ}{DQ}{data quality}
\newacronym{RDB}{RDB}{relational database}
\newacronym{DBMS}{DBMS}{database management system}
\newacronym{DB}{DB}{database}
\newacronym{FD}{FD}{Functional dependency}
\newacronym{OOV}{OOV}{out-of-vocabulary}
\newacronym{DBA}{DBA}{database administrator}
\newacronym{CFD}{CFD}{Conditional functional dependency}
\newacronym{DMV}{DMV}{Disguised missing value}
\newacronym{MVD}{MVD}{multivalued dependency}
\newcommand{\toterrornum}{35 }
\newcommand{\nullval}{$\bot$\xspace}
\newcommand{\NULL}{\data{NULL\xspace}}
\usepackage{xparse}
\usepackage{enumitem}

\usepackage{float}
\usepackage[table]{xcolor}

\usepackage{nicematrix}
\usepackage{subcaption}

\definecolor{lgray}{HTML}{F5F5F5} % F0F0F0
\usepackage{amsthm} % For definition
\usepackage{hyperref} % For definitionautoref
\newtheoremstyle{customdefinition} 
  {.5em} % space above the definition
  {.5em} % space below the definition
  {}        
  {1em} % indentation      
  {\normalfont} % not bold
  {.}      
  {.4em}   
  {}   

\newcounter{error}[section]
\renewcommand{\theerror}{\thesection.\arabic{error}} % 4.1, 4.2, 

\newtheorem{definition}{Definition}[section]

\theoremstyle{customdefinition}

\newcommand{\schema}[1]{\texttt{\small #1}\xspace}
\newcommand{\data}[1]{\texttt{\small #1}\xspace} 
\newcommand{\et}[1]{``#1''\xspace}
\newcommand{\true}{\textit{TRUE}\xspace}
\newcommand{\false}{\textit{FALSE}\xspace}

\NewDocumentCommand{\error}{o m +m +m +m}{%
  \par\addvspace{1em}%
  \refstepcounter{error}% <-- makes 'error' the current ref target
  \IfNoValueF{#1}{\label{#1}}% <-- attach label to the error counter
  \noindent\textbf{\theerror.}\ \textbf{#2.} #3\par
  \setcounter{definition}{\value{error}}%
  \begin{definition}#4\end{definition}%
  \par\noindent\textit{Example.} #5%
}

\acmJournal{CSUR}

\usepackage{lipsum}
\begin{document}

\title{A Catalog of Data Errors}
%\subtitle{Page limit: 35 pages incl. references}
% Alterantive:
%- Data Quality Through the Lens of Errors

\author{Divya Bhadauria}
\email{divya.bhadauria@hpi.de}
\affiliation{%
  \institution{Hasso Plattner Institute, University of Potsdam}
  \city{Potsdam}
  \country{Germany}
}

\author{Hazar Harmouch}
\email{h.harmouch@uva.nl}
\affiliation{%
  \institution{University of Amsterdam}
  \city{Amsterdam}
  \country{Netherland}
}

\author{Felix Naumann}
\email{felix.naumann@hpi.de}
\affiliation{%
  \institution{Hasso Plattner Institute, University of Potsdam}
  \city{Potsdam}
  \country{Germany}
}

\author{Divesh Srivastava}
\email{divesh@research.att.com}
\affiliation{%
  \institution{AT\&T Chief Data Office} 
  \city{Bedminster}
  \state{NJ}
  \country{USA}
}

\author{Lisa Ehrlinger}
\email{lisa.ehrlinger@hpi.de}
\affiliation{%
  \institution{Hasso Plattner Institute, University of Potsdam}
  \city{Potsdam}
  \country{Germany}
}

\renewcommand{\shortauthors}{Bhadauria et al.}

\begin{abstract}
 Data errors are widespread in real-world databases and severely impact downstream applications, such as machine learning pipelines
or business analytics reports. Causes of such errors are manifold and can arise during both the design phase and the operational phase
of a database. Some error types, such as missing values, duplicate tuples, or constraint violations, are widely recognized; others,
such as disguised missing values or word transpositions, remain underexplored. Existing attempts to define and classify errors in data
offer valuable but limited taxonomies, mostly informal and not covering the full range of error types. With the rise of AI, practitioners
must increasingly detect and correct statistical errors such as bias and outliers, which are rarely considered within existing error taxonomies. This catalog presents a comprehensive list of \toterrornum distinct error types, including both data errors (e.g., missing
values, duplicate tuples) and error indicators (e.g., outliers, bias) for tabular data, classified into three non-overlapping
categories: \emph{missing}, \emph{incorrect}, and \emph{redundant}. For each error type, we provide a formal definition and practical example, and
resolve terminological inconsistencies across related work. Our catalog enables researchers and practitioners to address various error
types and systematically implement error-specific detection and cleaning strategies in data quality tools.
\end{abstract}




%%\received{20 February 2007}
%%\received[revised]{12 March 2009}
%%\received[accepted]{5 June 2009}


% Undo comment for 3 lines
\maketitle

% \newpage
% \tableofcontents 
% \newpage

\section{Data Quality and Data Errors}
\label{sec:intro}

Data is ubiquitous; it has become an integral part of our daily lives and a key factor in data-driven decision-making across a wide range of industries and applications.
Many artificial intelligence (AI) applications, such as recommendation systems, weather forecasting, or resource planning, rely on machine learning (ML) pipelines to support a variety of tasks, including data pre-processing, ML model training, and result evaluation. 
Multiple studies~\cite{Colavito2024, Sedir2024, Bhadauria2024, Shen2024} have shown that the predictive performance of an ML model depends heavily on the quality of the training data. Given its impact, data quality (DQ) has also been widely researched and investigated for different domains, e.g., agriculture~\cite{Liakos2018, Elfatimi2024, Edita2024}, healthcare~\cite{Irvin2019, Yarmohammadtoosky2024}, and business~\cite{Xu2013, Deng2017, Petrovi2020, Youngjung2023, Wibisono2024}. To ground these efforts, DQ is typically defined as ``fitness for use''~\cite{Chrisman1984, Wang1996, Karr2006}, describing whether data is of sufficiently high quality for the intended use cases. 
Apart from technical challenges, poor DQ has been shown to cause significant financial loss in many sectors. 
According to the FMI Corporation report (2018)~\cite{Thomas2018}, poor DQ caused an annual monetary loss of USD 31.3~billion in the United States, USD 8.4~billion in Australia and New Zealand combined, and USD 10.2~billion in the United Kingdom, totaling USD 280~billion globally. 
Given the high impact of DQ across sectors such as business management, finance, and national growth,~\citet{Redman2023} argues that DQ poses a greater challenge than the development of ML models and software in the 21st century, highlighting the critical need for a deeper understanding of DQ.

There are many reasons for poor DQ, including issues during the data collection process (e.g., missing responses in forms) or issues that arise while working with the data (e.g., aggregating transaction values recorded in different currencies without conversion) -- both appearing during the \emph{operation phase} of an information system.
DQ issues can also appear during the \emph{design phase} of an information system, for example, a poorly designed schema can lead to redundancy.
In practice, these issues manifest in various ways, including data errors and error indicators, collectively referred to as \et{error types} in this catalog. 
A \emph{data error} is a mismatch between the intended and the actual representations of a real-world element (e.g., employee, department, or salary) in a database. 
In addition, we catalog \emph{error indicators}, which include logic-based patterns (e.g., FD violations) and statistical patterns (e.g., outliers, bias) that hint at errors in the data but require judgment to be defined as such.
While both data errors and error indicators can occur in any data storage system, this paper focuses on errors in relational databases (RDBs).
These errors can take many forms, such as missing values, incorrect formats, duplicate tuples, and spelling mistakes~\cite{Elmagarmid2007}. 
Such error types, along with many others, can have subjective interpretations, inherent uncertainty, unclear boundaries, and interdependency, making it difficult to define, detect, and correct them consistently.  

While traditional research on DQ has been studied through the lens of high-level DQ dimensions~\cite{Batini2009, Yuhan2024} (e.g., completeness, accuracy, timeliness), which define requirements for high-quality data primarily for managerial oversight, 
our catalog focuses on DQ from the low-level perspective of concrete errors in data (e.g., missing values, typos, outdated records) that can be directly detected in datasets. 
Both perspectives are valuable and complementary.
However, the low-level data-error perspective in this paper allows the investigation of what can be directly detected within a dataset and provides the fine-grained understanding required to quantify and interpret high-level DQ dimensions in practice~\cite{Papastergios_2025}.

To provide a clear definition and better understanding of the various error types in RDB data, researchers have previously proposed multiple taxonomies~\cite{RahmDo2000, Mueller2003, Kim2003, Oliveira2005, Ge2007ARO, Josko2016}. 
These taxonomies have classified specific error types, but there is no comprehensive list that consolidates the variety of these error types, includes the granularity levels at which they can be detected, along with formal definitions. 
Also, multiple uncommon error types (such as word transposition, misfielded values) remain underexplored, and no existing work distinguishes data errors from error indicators. 
Some works do not provide an explicit error type taxonomy but still distinguish data that is difficult to use for its intended purpose, such as missing values, inadequately defined values, measurements of questionable quality, or improperly integrated information from multiple sources~\cite{Lee2006}.
Understanding the various types of errors in data and their distinct characteristics can help researchers and practitioners identify and correct them.
For example, data science practitioners, such as data scientists, data engineers, and ML engineers~\cite{Hidalgo2024}, can use this catalog to (1)~implement validation checks for specific data errors and error indicators, (2)~plan data cleaning techniques for these error types, or (3)~identify under-researched error types that lack tool support for their detection.

\subsection{Contributions}
\label{subsection:contribution}
Our catalog approaches the field of DQ by cataloging data errors and error indicators, which characterize poor-quality data. 
The main contributions of this paper are: 
\begin{enumerate}
    \item A comprehensive catalog of \toterrornum data error types and error indicators, classified according to their manifestation into three categories: \emph{missing} data, \emph{incorrect} data, and \emph{redundant} data. 
    \item A formal definition and examples for each data error and error indicator. 
    \item A discussion on the connection between error types and errors in metadata, related data characteristics, as well as the use of this catalog in practice
    to improve error detection and cleaning.
    %\item A discussion on how to use these definitions in practice to improve error detection and cleaning
\end{enumerate}

To improve clarity, we applied two strategies to consolidate terminology across various error types. We started with simplifying overly complex or even misused labels with more intuitive names, such as renaming \et{Violation of Company and Government Regulations}~\cite{Ge2007ARO} to \et{Legal Rule Violations}. 
We then addressed and provided appropriate references for inconsistencies across taxonomies where error types are either described under multiple names or, conversely, where the same name is used for different errors.
For example, the term \et{contradiction} is used by Müller et al.~\cite{Mueller2003} for both \et{functional dependency violations} and \et{duplicate tuples}. In contrast, Rahm and Do~\cite{RahmDo2000} use this term only for duplicate tuples. 
For clarity and consistency, \et{contradictions} will not be treated as a separate error type in this catalog and will be mentioned under both \et{functional dependency violations} and \et{duplicate tuples}.

\subsection{Scope and Methodology}
This catalog provides a comprehensive discussion to \emph{understand} errors in data of RDBs.
We include data errors and error indicators that meet at least one of the following criteria: (1)~the data error (or error indicator) directly affects data usability and decision-making in a database, (2)~the data error (or error indicator) can be detected using systematic methods like rule-based systems, statistical algorithms, or ML models. 
For example, missing values meet both criteria, as they affect data usability and can be systematically detected using rules or ML algorithms~\cite{Zhou2024}. In contrast, semantically ambiguous data cannot be systematically detected but still satisfy the first criterion.
Our selection of each data error and error indicator in this catalog aims to provide practical value for researchers and practitioners who identify and correct these error types. 

The catalog focuses only on errors in data of RDBs and does not cover errors in other data modalities, such as graphs or text documents, which are interesting directions for future work. Further, we focus on \emph{operational} errors in base data and briefly discuss errors in metadata (appearing during the \emph{design phase}), such as schemata, constraints, or statistics, in~\autoref{sec:metadata}.
We also do not look over the vast areas of methods and tools to detect and clean data errors and error indicators, but discuss their connection to this catalog in~\autoref{sec:errormanagment}.

The basis of our catalog is five existing taxonomies~\cite{RahmDo2000, Kim2003, Oliveira2005, Ge2007ARO, Josko2016} on data errors, which we summarize in~\autoref{sec:related-work}. We first screened the existing survey papers on data errors for inclusion and assessed the quality of these taxonomies, as also suggested by~\citet{Templier2015}.
While these taxonomies are a good basis for understanding data errors, they do not provide a comprehensive list of the wide spectrum of error types. We address this research gap by creating a joint list from these taxonomies and extending the catalog in two directions: 
(i) we identified subtypes and variants of the error types discussed in these works, and (ii) we added recently emerged error types.
The fact that a few error types were not included or were only partially discussed in previous taxonomies (such as disguised missing values, invalid values, and out-of-vocabulary values) further underscores the need for a refreshed, comprehensive errors catalog.
We assembled our error catalog by conducting in-depth examinations of each data error type and reviewing research on those error types.
Following this initial investigation, we clustered all error types and indicators into three mutually exclusive categories: \emph{missing} data, \emph{incorrect} data, and \emph{redundant} data, based on how the errors are manifested in a dataset.
While we recognize that other specialized error types exist, including them would add length without improving the comprehensiveness and value of this catalog.

\subsection{Paper Outline}
\autoref{sec:preliminaries} introduces a formal notation to describe error types, a running example on which our examples are based, as well as a description of our classification dimensions.
In the following three sections, we define various types of errors: missing data in \autoref{sec:missingdata}, incorrect data in \autoref{sec:incorrectdata}, and redundant data in \autoref{sec:redundantdata}. \autoref{sec:metadata} discusses errors in metadata, 
\autoref{sec:related-data-characteristics} covers related data characteristics which impact data errors and error indicators but are not errors themselves, and \autoref{sec:errormanagment} summarizes methods and tools to detect and correct errors in data. 
In \autoref{sec:related-work}, we review related work used for this catalog and conclude in \autoref{sec:conclusion}.
%\newpage

\section{Preliminaries and Error Type Framework}
\label{sec:preliminaries}
% \glsreset{RDB}
% \glsreset{OOV}
% \glsreset{FD}
% \glsreset{DQ}

In this section, we first present a database excerpt as a basis for all examples mentioned in this catalog in \autoref{subsection:running-example}. Then, we introduce the formal notations that are used to define all error types in \autoref{subsection:Preliminaries} and explain the classification dimensions used to organize them in \autoref{sec:classification}. To show the mismatch between a value in the real-world and its database representation, we distinguish between the \emph{real-world element} and its \emph{database representation}. 
A real-world element refers to an entity or relationship as it exists in reality and is described by finite, approximate values. These values are stored as representations in the database. An error-free database representation should exactly match its intended real-world counterpart.

\subsection{Running Example}
\label{subsection:running-example}
\begin{figure}[ht]
    \centering
    \includegraphics[width=0.85\textwidth, trim=50 233 110 181, clip]{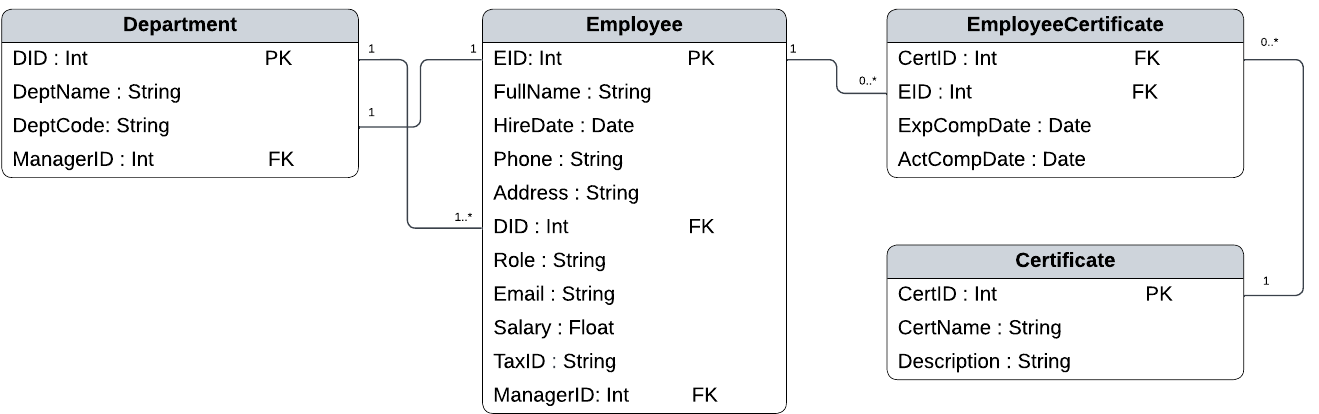}
    \caption{Entity-relationship diagram for the running example: \schema{Employment} database.}
    \Description{}
    \label{fig:ERDiagram}
\end{figure}

This section introduces a running example based on the \schema{Employment} database, which is used throughout this catalog to illustrate each data error and error indicator. The example database stores information about employees, their affiliations, and certificates for further training. The entity-relationship diagram of the \schema{Employment} database is as shown in \autoref{fig:ERDiagram}, and data in each relation are shown in \autoref{fig:ExampleInstances}, where the grayed out tuples indicate the correct corresponding real-world values of tuples that have incorrect values.
The \schema{Employee} relation references the \schema{DID} attribute of the \schema{Department} relation, and its \schema{MID} attribute (Manager ID of an attribute) references \schema{EID}, making \schema{Employee} a self-referencing relation. 
The \schema{ManagerID} attribute in the \schema{Department} relation references the \schema{EID} attribute in the \schema{Employee} relation, establishing a relationship in which managers can also be employees. 
This design introduces a cyclic dependency between \schema{Employee} and \schema{Department} relations.
As a result, the schema does not guarantee a hierarchical structure and may contain cycles. The \schema{EmployeeCertificate} relation references \schema{CertID} and \schema{EID} from the \schema{Certificate} and \schema{Employee} relations, respectively. 

\begin{figure}[htbp]
\centering

% ---------- Example instances ----------
\begin{subfigure}{\textwidth}
\centering
\begin{adjustbox}{width=\textwidth}
\begin{tabular}{c|c|l|l|c|c|c|c|c|c|c|c}
\hline
\toprule
\multicolumn{12}{c}{\textbf{Employee}} \\
\midrule
& EID & FullName & HireDate & Phone & Address & DID & Role & Email & Salary & TaxID & MID\\
\midrule
e1 & 101 & John Taylor & $\bot$ & $\bot$ & Green Street, California, 90001 & 10 & Software Engineer & john.taylor@cmpdom.com & 40,000 & 56 897 086 787 &  102
\\
e2 & 102 & Eric J. Smith  & $\bot$ & $\bot$ & CA 91 Street, California, 99801 & 40 & Mgr. & eric.j.smith@cmp.com & 67,000 & 12 345 678 901 & 101
\\
e3 & 103 & Jessica, Jones  & 14/03/2023 & $\bot$ & George Street 48, California, 99456 & 20 & Human Resource & jessica.jones@cmp.com & -65,000 & 12 355 678 651 & $\bot$
\\
e4 & 104 & Nancy 34ller & 20.08.2024 & $\bot$ & Linkon Avenue, NY & 30 & Security & sera.müller@ & 50k & 98 798 798 798 & 107
\\
e5 & 105 & Sara Moller & 01/08/2024 & $\bot$ & Queens Villa 59, CA & 10 & Software Engineer & sera.müller@itcmp.com & 52,600 & 90 790 790 790 & 102
\\
e6 & 106 & Bond, James & 05/04/2022 & $\bot$ & Lincoln Avenue, CA & 10 & Software Engineer & bond.james@itcmp.com & 68,600 & 90 709 709 708 & 102 \\
e7 & 107 & Marcus D’Angelo & Angelo 01/01/2023 & $\bot$ & London Street, CA & 10 & Head of Security & marcus.d@itcmp.com & 205,000 & 80 977 797 708 & $\bot$ \\
e8 & 103 & Jessica J.  & 14/03/2023 & $\bot$ & George Street 48, California, 99456 & 20 & Human Resource & jessica.j@cmp.com & -65,000 & 12 355 678 651 & $\bot$
\\
e9 & 101 & J. Taylor & $\bot$ & $\bot$ & Green Street, California, 90001 & 10 & Software Engineer & john.taylor@cmpdom.com & 40,000 & 56 897 086 787 & 102
\\
 \midrule
\textcolor{gray}{M(e1)} & \textcolor{gray}{101} & \textcolor{gray}{John Taylor} & \textcolor{gray}{01/04/2023} & \textcolor{gray}{+1 (478) 314-1657} & \textcolor{gray}{78 Green Street, California, 90001} & \textcolor{gray}{10} & \textcolor{gray}{Software Engineer} & \textcolor{gray}{john.taylor@cmpdom.com} & \textcolor{gray}{40,000} & \textcolor{gray}{56 897 086 787} & \textcolor{gray}{102}
    \\
\textcolor{gray}{M(e2)} & \textcolor{gray}{102} & \textcolor{gray}{Eric J. Smith}  & \textcolor{gray}{01/01/2024} & \textcolor{gray}{+1 (549) 407-5104} & \textcolor{gray}{91 CA Street, California, 99801} & \textcolor{gray}{40} & \textcolor{gray}{Manager} & \textcolor{gray}{eric.j.smith@cmpdom.com} & \textcolor{gray}{67,000} & \textcolor{gray}{92 765 876 554} & \textcolor{gray}{$bot$}
    \\
\textcolor{gray}{M(e3), M(e8)} & \textcolor{gray}{103} & \textcolor{gray}{Jessica, Jones}  & \textcolor{gray}{14/03/2023} & \textcolor{gray}{+1 (334) 248-3147} & \textcolor{gray}{48 George Street, California, 99456} & \textcolor{gray}{20} & \textcolor{gray}{Human Resource} & \textcolor{gray}{jessica.jones@cmpdom.com} & \textcolor{gray}{65,000} & \textcolor{gray}{92 355 678 651} & \textcolor{gray}{$bot$}
    \\
\textcolor{gray}{M(e4)} & \textcolor{gray}{104} & \textcolor{gray}{Nancy Müller} & \textcolor{gray}{20.08.2024} & \textcolor{gray}{+1 (765) 327-4012} & \textcolor{gray}{66  Lincoln Avenue, NY} & \textcolor{gray}{30} & \textcolor{gray}{Security} & \textcolor{gray}{sera.müller@cmpdom.com} & \textcolor{gray}{50,000} & \textcolor{gray}{98 798 798 798} & \textcolor{gray}{107}
    \\
\textcolor{gray}{M(e5)} & \textcolor{gray}{105} & \textcolor{gray}{Sara Müller} & \textcolor{gray}{01/08/2024} & \textcolor{gray}{+1 (765) 347-2451} & \textcolor{gray}{59 Queens Villa, CA} & \textcolor{gray}{10} & \textcolor{gray}{Software Engineer} & \textcolor{gray}{sera.müller@itcmpdom.com} & \textcolor{gray}{52,600} & \textcolor{gray}{90 790 790 790} & \textcolor{gray}{102}
    \\
\textcolor{gray}{M(e6)} & \textcolor{gray}{106} & \textcolor{gray}{Bond, James} & \textcolor{gray}{05/04/2022} & \textcolor{gray}{+1 (903) 468-7249} & \textcolor{gray}{15 Lincoln Avenue, CA} & \textcolor{gray}{10} & \textcolor{gray}{Software Engineer} & \textcolor{gray}{bond.james@itcmpdom.com} & \textcolor{gray}{68,600} & \textcolor{gray}{90 709 709 708} & \textcolor{gray}{102}
    \\
\textcolor{gray}{M(e7)} & \textcolor{gray}{107} & \textcolor{gray}{Marcus D’Angelo} & \textcolor{gray}{01/01/2023} & \textcolor{gray}{+1 (934) 604-7171} & \textcolor{gray}{29 London Street, CA} & \textcolor{gray}{30} & \textcolor{gray}{Head of Security} & \textcolor{gray}{marcus.dangelo@itcmpdom.com} & \textcolor{gray}{205,000} & \textcolor{gray}{80 977 797 708} & \textcolor{gray}{$bot$}
    \\
\textcolor{gray}{M(e9)} & \textcolor{gray}{101} & \textcolor{gray}{J. Taylor} & \textcolor{gray}{01/04/2023} & \textcolor{gray}{+1 (802) 710-9663} & \textcolor{gray}{31 Green Street, California, 90001} & \textcolor{gray}{10} & \textcolor{gray}{Software Engineer} & \textcolor{gray}{john.taylor@cmpdom.com} & \textcolor{gray}{40,000} & \textcolor{gray}{56 897 086 787} & \textcolor{gray}{102}
    \\
\hline
\end{tabular}
\end{adjustbox}
%\caption{\schema{Employee} relation}
\bigskip
\end{subfigure}

\noindent
\subcaptionbox*{\label{tab:certificate}}{%
  \begin{adjustbox}{width=.6\textwidth}
  \begin{tabular}{c|c|c|c}
  \toprule
  \multicolumn{4}{c}{\textbf{Certificate}} \\
  \midrule
  & CertID & CertName & Description \\
  \midrule
  c1 & S101 & SAP HANA Certification & SAP HANA Certificate \\
  c2 & C\_HANATEC\_18 & SAP HANA 2.0 SPS05 & Technology associate exam for SAP HANA 2.0 \\
  c3 & C-HANATEC-18 & SAP HANA 2.0 Technology Associate & Certification for SAP HANA 2.0 administrators \\
  c4 & MS\_AZ\_900 & Microsoft Azure Fundamentals & Introductory cloud certification from Microsoft \\
  \bottomrule
  \end{tabular}
  \end{adjustbox}
}
\hfill
\subcaptionbox*{\label{tab:EmployeeCertificate}}{%
  \begin{adjustbox}{width=.36\textwidth}
  \begin{tabular}{c|c|c|l|l}
      \toprule
      \multicolumn{5}{c}{\textbf{EmployeeCertificate}} \\
      \midrule
      & CertID & EID & ExpCompDate & ActCompDate \\
      \hline
      ec1 & S101 & 101 & 12/08/2025 & 11/08/2025 
      \\
      ec2 & S101 & $\bot$ & 12/02/2024 & $\bot$ 
      \\
      ec3 & C\_HANATEC\_18 & 104 & 31/12/2023 & 12/12/2023 
      \\
      \midrule
      \textcolor{gray}{M(ec2)} & \textcolor{gray}{14} & \textcolor{gray}{105} & \textcolor{gray}{12/02/2024} & \textcolor{gray}{12/02/2024} 
      \\
      \bottomrule
  \end{tabular}
  \end{adjustbox}
}
\centering
% ---------- Department ----------
\begin{subfigure}{.30\textwidth}
\centering
\begin{adjustbox}{width=\textwidth}
    \begin{tabular}{c|c|c|c|c}
    \toprule
        \multicolumn{4}{c}{\textbf{Department}} \\
        \midrule
        & DID & DeptName & DeptCode & ManagerID \\
        \midrule
        d1 & 20 & Human Resource & HR & 103 \\
        d2 & 30 & Security & ST &$\bot$ \\
        d3 & 10 & IT & IT &102 \\
        d4 & 40 & Management & MT & $\bot$ \\
        d5 & 50 & HR & HR & 103 \\
        d6 & 20 & Human R. & HR & 105 \\
        d7 & 60 & Marketing & MR & $\bot$ \\
        \midrule
        \textcolor{gray}{M(d2)} & \textcolor{gray}{30} & \textcolor{gray}{Security} & \textcolor{gray}{ST} & \textcolor{gray}{104}
        \\
        \textcolor{gray}{M(d4)} & \textcolor{gray}{40} & \textcolor{gray}{Management} & \textcolor{gray}{MT} & \textcolor{gray}{102}
        \\
        \bottomrule
    \end{tabular}
\end{adjustbox}
\caption*{\label{tab:Department}}
\end{subfigure}
\caption{Example instances of \schema{Employee}, \schema{Certificate}, \schema{Employeecertificate}, and \schema{Department} relations. 
Grayed-out tuples indicate the corresponding correct values of each attribute for a real-world element, indicating that the database values in those tuples are incorrect. For instance, if a tuple $t$ is incorrect, we represent the real-world values in a greyed-out tuple $t'$.}
\label{fig:ExampleInstances}
\Description{}
\end{figure}

\subsection{Preliminaries}
\label{subsection:Preliminaries}
In this section, we introduce notations to describe 
(1)~data in a database and 
(2)~the mapping of these data to the corresponding real-world elements as a basis to formally describe data errors.

A database is denoted by $db$, defined as a set of relations $R$  
%= \{r_1, r_2, \ldots, r_o\}$
, where each relation $r_i \in R$ has a set of attributes $A = \{a_{1}, a_{2}, \ldots, a_{m}\}$
and tuples $T = \{t_{1}, t_{2}, \ldots, t_{n}\}$. 
We denote a database element by $e$, and it is expressed using dot-notation (such as $db.r.t.a$, $db.r.a$, or $t.a$). For instance, $db.r.t.a$ refers to the value of attribute $a$ in tuple $t$ within relation $r$ of database $db$. 
When the context is clear, we omit the leading prefixes, and when the identity of the individual component is irrelevant, we omit trailing suffixes. 
For example, we omit the attribute reference (i.e., $a$) when referring to an entire tuple (i.e., $r.t$) and the tuple reference (i.e., $t$) when referring to an attribute (i.e., $r.a$) within a relation $r$. Similarly, we may omit the reference to relation $r$ when it is clear from context. 
We refer to stored values in a database element using a value accessor ($e.v$). For example, $t.a.v$ denotes the value stored in attribute $a$ in tuple $t$, $r.t.v$ denotes the list of values of all attributes in tuple $t$, and $r.a.v$ denotes the list of all values in attribute $a$ across all tuples of relation $r$.
We also omit the notion of time by default, but refer to a specific point in time with $\tau$ in superscript when discussing temporal data, such as $t^\tau$ denoting a tuple $t$ observed at time~$\tau$.
%-- the \emph{extension}~\cite{Chopra2010} -- 

To define mismatches between a database representation and the value of a corresponding real-world element, we define a mapping function $M(e)$ that maps a database element $e$ to its corresponding real-world representation. 
% This function maps a database element $e$ at the corresponding level of granularity to its real-world origins.
For instance, $M(t.a)$ returns the expected representation of a real-world entity associated with the attribute $a$ in tuple $t$.
Using this mapping function, a value-level mismatch can be written as $t.a.v \neq M(t.a)$. 
Since $M(e)$ only maps existing data, it cannot capture entities that are missing at any level of granularity. To define missing tuples, we introduce an additional mapping function $M^*(r)$ (specific to relations) that returns the set of all real-world entities that relation $r$ is expected to represent, given its schema. 

We represent explicit missing values in the database using the $\bot$ symbol, and to represent disguised missing values (such as ``Unknown'' or $-99$), we use a mapping function $M(\bot)$ that represents \data{NULL} in the real world.
To define a partially empty tuple/attribute (\autoref{def:partial-empty-tuple}) and empty attribute (\autoref{def:empty-attribute}), we define a count function $\textit{Count}_{MV}(e.v)$, which counts the number of both explicit and disguised missing values at the given database element level. 
For the definition of out-of-vocabulary (OOV) errors in \autoref{err:OOV-word} and \autoref{err:spelling}, we define a $OOV(w, V_a)$ function, which determines whether a word, denoted by $w$, is in the vocabulary of the target language, denoted by $V_a$. The $OOV(w, V_a)$ function returns \true if and only if the word $w$ is not present in the target vocabulary. 
Here, $w$ may also include alphanumeric characters. When a data value contains a single word, $w$ is the complete value; otherwise, $w$ can be extracted through preprocessing from multi-word text (e.g., a value in \schema{Certificates.Description} attribute). 
We exclude explicit missing values in the $OOV(w, V_a)$ function, as they represent completeness issues rather than a vocabulary-related error. 
% Unit(a) is used in incorrect unit definition
We define a $\textit{Unit}(a)$ function which returns the measurement unit associated with an attribute $a$, assuming that attribute 
$a$ is unit-homogeneous, i.e., all its values should be expressed in a common measurement unit. Any deviation from this assumption will indicate an inconsistency.
% Syntax(t.a) is used in syntax violation definition
We also define a $\textit{Syntax}(t.a.v)$ function that returns the syntax of the value of database element $t.a$.

To cover various constraint violations, we define a function, $\textit{DependencyCheck}(A_{lhs}, A_{rhs}, c)$ that returns \true if a dependency relationship holds between a set of attributes on the left-hand side (denoted by $A_{lhs}$) and a list of attributes on the right-hand side (denoted by $A_{rhs}$) in relation $r$ under a given set of optional conditions, denoted by $c$. 
Here, the left-hand side and right-hand side denote the determinant and dependent attributes of a dependency, respectively, and carry no positional meaning in the schema.

\begin{figure}
    \centering
    \includegraphics[height=0.95\textheight]
    {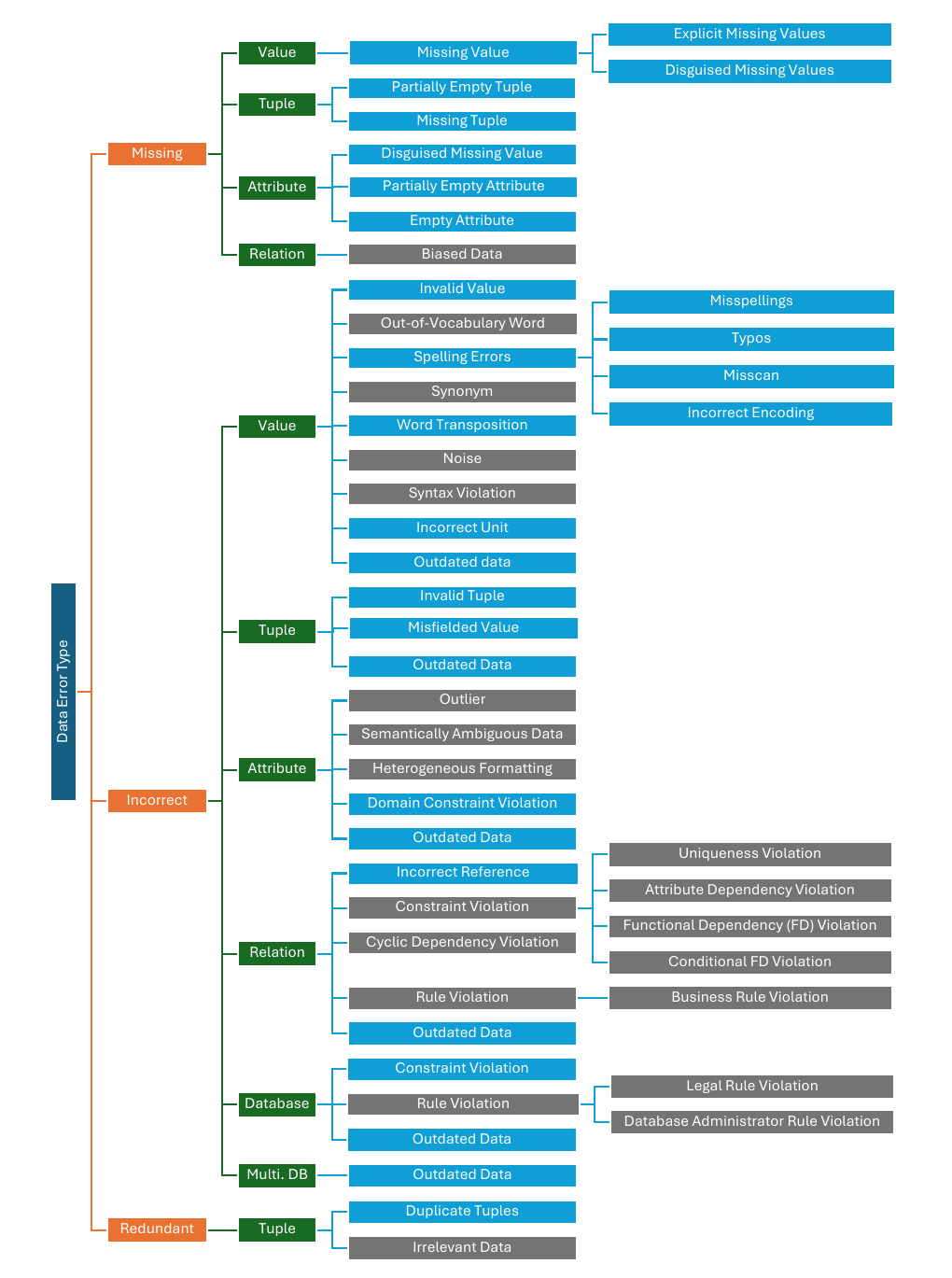}
    \caption{A hierarchy of data error types (indicated with blue background) and error indicators (indicated with gray background).}
    \Description{}
    \label{fig:errorhierarchy}
\end{figure}

\subsection{Classification Dimensions}
\label{sec:classification}
Given the many perspectives on errors in data of a relational database (RDB), the literature has identified various means to group or classify them. Typical classifications include semantic vs.\ syntactic errors~\cite{Mueller2003}, single vs.\ multiple source errors~\cite{RahmDo2000}, user vs. data perspective~\cite{Ge2007ARO}, and context-independent vs.\ -dependent~\cite{Ge2007ARO}, but how a specific error type should be assigned is not always clear. For example, it might not be straightforward to argue whether a functional dependency violation is of \emph{syntactic} or \emph{semantic} nature, and such distinctions require a clear understanding before categorization. 
We argue that such a classification remains helpful for gaining an overview of the broad and heterogeneous nature of data errors and error indicators. 
This section explains the dimensions used to classify error types as shown in~\autoref{tab:errorclassification}. The data errors are highlighted in light gray, and the error indicators in white. We also show how error types are organized into these three categories at different granularity levels in \autoref{fig:errorhierarchy}, along with which specific error types belong to each broader category (e.g., one of the broad categories is ``missing value'' and subtypes under this are explicit missing values and disguised missing values). The error indicators are highlighted in gray, and data errors are highlighted in blue.

\textbf{Data error manifestation}
\label{subsubsec:manifest}
is our primary classification dimension, in which error types are classified by how they manifest in the data, ensuring that errors in each category are mutually exclusive. The three categories are \emph{missing}, \emph{incorrect}, and \emph{redundant}. Error types in the missing category occur when required data are absent in the database, even though the corresponding real-world entity actually has a valid value. When the real-world entity lacks such a value, it indicates a schema or modeling issue rather than a missing data error. Error types in the incorrect section include issues that do not accurately represent the intended real-world information. Error types in the redundant section include unnecessary repetition or data duplication.

\textbf{Data error granularity}
\label{subsubsec:granularity}
is another way to classify error types~\cite{Oliveira2005}, i.e., from value-, tuple-, or attribute-level over table-level to DB-level data errors. 
Note that the granularity at which an error type is introduced may differ from that at which it can be detected. 
For example, a violation of a functional dependency (cf.~\autoref{err:fd-violation}) can only be detected at the table level by regarding multiple attributes and multiple tuples at the same time, whereas the respective dependency can be violated solely by inserting a single tuple with an incorrect attribute value.
This paper uses the granularity level at which the errors are detected as one approach for their classification.

\textbf{Data error context}
\label{subsubsec:context}
distinguishes error types as \emph{syntactic} or \emph{semantic}~\cite{Ge2007ARO, Fan2012}.
Syntactic category includes error types where the data format deviates from predefined rules, e.g., a date \data{01.09.2024} is stored in the attribute \schema{HireDate} which expects ``\schema{dd/mm/yyyy}'' format instead. 
Semantic category includes error types that do not represent the intended meaning or real-world element, e.g., assigning the same employee ID to multiple employees in an organization results in incorrect record identification or linkage.

\begin{table}
    \centering
    \begin{adjustbox}{width=\textwidth}
    \begin{tabular}{c|ccc|cccccc|cc}
    \toprule
    Error Manifestation & \multicolumn{3}{l|}{Error Type} & \multicolumn{6}{l|}{Data Granularity} & \multicolumn{2}{l}{Context} \\ 
        & & & & \rotatebox{90}{Value} & \rotatebox{90}{Tuple} & \rotatebox{90}{Attribute} & \rotatebox{90}{Relation} & \rotatebox{90}{DB} & \rotatebox{90}{Mult. DBs} & \rotatebox{90}{Syntactic} & \rotatebox{90}{Semantic} \\
        \midrule 
         \multirow{6}{*}{Missing} & \multicolumn{3}{l|}{\cellcolor{lgray}\hyperref[err:missing-val]{Missing Value}~\cite{RahmDo2000, Mueller2003, Oliveira2005, Ilyas2022}} & \textsf{X} & & & & & & & \textsf{X}  \\
          & \multicolumn{3}{l|}{\cellcolor{lgray}\hyperref[err:disguised-missing-val]{Disguised Missing Value}~\cite{Pearson2006, Qahtan2018}} & \textsf{X} & & & & & & \textsf{X} & \textsf{X} \\
          & \multicolumn{3}{l|}{\cellcolor{lgray}\hyperref[err:partial-empty-tuple]{Partial-Empty Tuple/Attribute}~\cite{Oliveira2005}} & & \textsf{X} &  \textsf{X} & & & & & \textsf{X} \\
          & \multicolumn{3}{l|}{\hyperref[err:missing-tuple]{\cellcolor{lgray}Missing Tuple}~\cite{Josko2016, Mueller2003}} & & \textsf{X} & & & & & & \textsf{X}  \\
          & \multicolumn{3}{l|}{\cellcolor{lgray}\hyperref[err:empty-attribute]{Empty Attribute}~\cite{Grahne2018}} & & & \textsf{X} & & & & \textsf{X} & \textsf{X} \\
          & \multicolumn{3}{l|}{\hyperref[err:biased-data]{Biased Data}~\cite{Lee2006, Emilio2024, Kurapati2025}} & & & & \textsf{X} & & & & \textsf{X} \\
      \midrule
         \multirow{30}{*}{Incorrect}
            %& \multicolumn{2}{l|}{Incorrect Value~\cite{Ge2007ARO, Oliveira2005, Josko2016}} & \textsf{X} & & & & & & & \textsf{X} & \textsf{X} & \\
            & \multicolumn{3}{l|}{\hyperref[err:invalid-val-tup]{Invalid Value/Tuple}~\cite{Mueller2003, RahmDo2000}} & \textsf{X} & \textsf{X} & & & & & & \textsf{X} \\
            \cline{2-4}
            & \multirow{7}{*}{%
            $\left.\vphantom{\begin{array}{l}1\\2\\3\\4\\5\\6\end{array}}\vcenter{\hbox{\rotatebox{90}{\strut Textual}}}\right\{$} & \multicolumn{2}{l|}{\hyperref[err:OOV-word]{Out-of-Vocabulary Word}~\cite{Chen2019}} &  \textsf{X} & &  &  &  & & & \textsf{X} \\
            & & \multicolumn{2}{l|}{\hyperref[err:spelling]{\cellcolor{lgray}Misspelling}~\cite{RahmDo2000, Kim2003, Oliveira2005}} & \textsf{X} & & & & & & & \textsf{X} \\
            & & \multicolumn{2}{l|}{\hyperref[err:spelling]{\cellcolor{lgray}Typo}~\cite{Shah2020, Cox2017}} & \textsf{X} & & & & & & & \textsf{X} \\
            & & \multicolumn{2}{l|}{\hyperref[err:spelling]{\cellcolor{lgray}Misscan}~\cite{Nguyen2019,Kissos2016}} & \textsf{X} & &  &  &  & & & \textsf{X} \\
            & & \multicolumn{2}{l|}{\hyperref[err:spelling]{\cellcolor{lgray}Incorrect Encoding}~\cite{Kim2003}} & \textsf{X} & & & & & & \textsf{X} & \\
            & & \multicolumn{2}{l|}{\hyperref[err:synonyms]{Synonym}~\cite{Josko2016, Mu2023}}& \textsf{X} & & & & & & & \textsf{X} \\
            & & \multicolumn{2}{l|}{\hyperref[err:word-transposition]{\cellcolor{lgray}Word Transposition}~\cite{RahmDo2000, Kim2003}} & \textsf{X} & & & & & & & \textsf{X} \\
            \cline{2-4}
            & \multirow{8}{*}{$\left.
            \vphantom{\begin{array}{l}1\\2\\3\\4\\5\\6\\7\end{array}}
            \vcenter{\hbox{\rotatebox{90}{\strut Nonconformant}}}
            \right\{$} & \multicolumn{2}{l|}{\hyperref[err:misfielded-val]{\cellcolor{lgray}Misfielded Value}~\cite{RahmDo2000}}  & & \textsf{X} & & & & & & \textsf{X} \\
            & & \multicolumn{2}{l|}{\hyperref[err:noise]{Noise}~\cite{Subramaniam2009, Winkler2003}} & \textsf{X} & & & & & & & \\
            & & \multicolumn{2}{l|}{\hyperref[err:semantic-amb]{Semantically Ambiguous Data}~\cite{Kim2003, Oliveira2005}} & & & \textsf{X} & & & & & \textsf{X} \\
            & & \multicolumn{2}{l|}{\hyperref[err:outlier]{Outlier}~\cite{Ilyas2022}}  & & & \textsf{X} & & & & & \\
            & & \multicolumn{2}{l|}{\hyperref[err:syntax-viol]{Syntax Violation}~\cite{Ge2007ARO, Oliveira2005}} & \textsf{X} & & & & & & \textsf{X} &\\
            & & \multicolumn{2}{l|}{\hyperref[err:heterogeneous-formatting]{Heterogeneous Formatting}~\cite{Oliveira2005}} &  & & \textsf{X} & & & & \textsf{X} & \\
            & & \multicolumn{2}{l|}{\hyperref[err:incorrect-unit]{\cellcolor{lgray}Incorrect Unit}~\cite{Kim2003}}  & \textsf{X} & & & & & & \textsf{X} & \\
            % & & \multicolumn{2}{l|}{Heterogeneous Unit~\cite{Oliveira2005, Josko2016}}  & & & \textsf{X} & & & & \textsf{X} & \\
            & & \multicolumn{2}{l|}{\hyperref[err:incorrect-ref]{\cellcolor{lgray}Incorrect Reference}~\cite{Josko2016, Oliveira2005}}  & & & & \textsf{X} & & & & \textsf{X} \\
            \cline{2-4}
            & \multirow{10}{*}{$\left.
            \vphantom{\begin{array}{l}1\\2\\3\\4\\5\\6\\7\\8\\9\end{array}}
            \vcenter{\hbox{\rotatebox{90}{\strut Rule Violation}}}
            \right\{$} & \multicolumn{2}{l|}{\hyperref[err:const-violation]{Constraint Violation}~\cite{ Mueller2003, Ge2007ARO, Decker2011}} & & & & \textsf{X} & \textsf{X} & & & \textsf{X} \\
            & & \multicolumn{2}{l|}{\hyperref[err:domain-const-violation]{Domain Constraint Violation}~\cite{Josko2016, Ge2007ARO, Oliveira2005}} & & & \textsf{X} & & & & & \textsf{X} \\
            & & \multicolumn{2}{l|}{\hyperref[err:uniqueness-violation]{Uniqueness Violation}~\cite{Ge2007ARO, Oliveira2005, RahmDo2000}} &  & & & \textsf{X} & & & \textsf{X} & \\
            & & \multicolumn{2}{l|}{\hyperref[err:attr-dep-violation]{Attribute Dependency Violation}~\cite{RahmDo2000}} & & & & \textsf{X} & & & & \textsf{X} \\
            & & \multicolumn{2}{l|}{\hyperref[err:fd-violation]{Functional Dependency (FD) Violation}~\cite{Josko2016, Oliveira2005, Ge2007ARO}} & & & & \textsf{X} & & & & \textsf{X} \\
            & & \multicolumn{2}{l|}{\hyperref[err:cfd-violation]{Conditional FD Violation}~\cite{Josko2016}} & & & & \textsf{X} & & & & \textsf{X} \\
            & & \multicolumn{2}{l|}{\hyperref[err:cyclic-dep-violation]{Cyclic Dependency Violation}~\cite{Oliveira2005}} & & & & \textsf{X} & & & & \textsf{X} \\
            & & \multicolumn{2}{l|}{\hyperref[err:rule-violations]{Business Rule Violation}~\cite{Ge2007ARO, Oliveira2005}}  & & & & \textsf{X} & & & & \textsf{X} \\
            & & \multicolumn{2}{l|}{\hyperref[err:rule-violations]{Database Administrator Rule Violation}~\cite{Ge2007ARO}} & & & & & \textsf{X} & & & \textsf{X} \\ 
            & & \multicolumn{2}{l|}{\hyperref[err:rule-violations]{Legal Rule Violation}~\cite{Ge2007ARO}}  & & & & & \textsf{X} & & & \textsf{X}
            \\
            \cline{2-4}
            & \multicolumn{3}{l|}{\hyperref[err:outdated-data]{\cellcolor{lgray}Outdated Data}~\cite{Ge2007ARO}} & \textsf{X} & \textsf{X} & \textsf{X} & \textsf{X} & \textsf{X} & \textsf{X} & & \textsf{X} \\
     \midrule
         \multirow{2}{*}{Redundant} & \multicolumn{3}{l|}{\hyperref[err:duplicate-tuples]{\cellcolor{lgray}Duplicate Tuples}~\cite{Ge2007ARO, RahmDo2000, Mueller2003,Oliveira2005,Josko2016}} & & \textsf{X} & & & & & & \textsf{X} \\
            & \multicolumn{3}{l|}{\hyperref[err:irrelevant-row]{Irrelevant Data}~\cite{Ge2007ARO}} & & \textsf{X} & & & & & & \textsf{X} \\
    \bottomrule
    \end{tabular}
    \end{adjustbox}
    \caption{Classification of error types based on how they manifest in data as main category, and two other possible classifications, granularity, and context. Data errors are highlighted in light gray; error indicators in white.}
    \label{tab:errorclassification}
\end{table}

%\medskip
With these classifications in mind, we focus on the three different error manifestation types (missing, incorrect, and redundant) and, in the following sections, define and explain the various error types within their respective classes.

\section{Missing Data}
\label{sec:missingdata}
\glsreset{RDB}
\glsreset{OOV}
\glsreset{FD}
\glsreset{DQ}
% verify the new sentence for these citations: ~\cite{RahmDo2000, Alsufyani2024}
Data in a relational database (RDB) are considered \emph{missing} when the database representation of a real-world element is absent from the database.
Missing data in a mandatory attribute is always considered an error, as it violates the \schema{NOT NULL} schema constraint. 
On the other hand, missing data in a non-mandatory attribute can result in errors if it negatively impacts downstream tasks or leads to incorrect interpretations~\cite{RahmDo2000, Alsufyani2024}.
Missing data are widespread in RDBs, particularly in scenarios involving data integration, data cleaning, and data exchange~\cite{Antova2007}.
It can be caused by delays or omissions in data entry, the use of self-reported data~\cite{Peng2023}, data loss due to system downtime or bottlenecks, and faulty sensors or devices that fail to capture measurements accurately. 
Although the default representation for missing data in RDBs is \NULL, placeholders like ``\data{Unknown}'', ``\data{NA}'', or ``\data{Missing}'' for string attributes, \data{999} or \data{-1} for numeric attributes, or \data{000-000-0000} for phone numbers (for USA) can also represent missing data. We use $\bot$ to represent \NULL values in this catalog.

The literature also categorizes missing data according to the mechanism causing them~\cite{Little_1983, Peng2023}: data can be missing completely at random (MCAR), where the missingness occurs independently of any attribute, including the affected one; missing at random (MAR) where the missingness depends on the value of at least one attribute other than the affected one; and missing not at random (MNAR), where the missingness depends on the affected attribute itself and may also depend on the other attributes. 
% These categorizations take a statistical approach to explain the distribution and the reason behind the missing data. 
In addition, missing data can be categorized according to different patterns of missingness, such as univariate, where only a single attribute has missing values; monotone, where missing values follow a step-like pattern; or general, where missing values can be found in any attribute values~\cite{Emmanuel2021,Bechny2021}.
Understanding the different types of missingness patterns and the mechanisms that lead to missing data can help detect and clean them more efficiently, as different occurrence patterns and distributions may require different handling approaches. In addition, knowledge of these mechanisms can help prevent missing data in the first place, for instance, by enforcing integrity constraints such as \schema{NOT NULL} on an attribute.
In the following subsections, we discuss the errors in the missing category at various granularity levels. 
%This categorization can be a first step for quickly detecting data errors before applying complex statistical analyses of patterns and mechanism.

%For one table (including its cells, rows, and columns), 
\error[err:missing-val]
{(Explicit) Missing value}
% ----- Description ------------------------------------------------------------
{
Missing values can occur in both mandatory and non-mandatory
attributes within a relation and by default are represented as \NULL in an RDB. 
While the absence of a value in a mandatory attribute is clearly an error as it violates the \schema{NOT NULL} constraint~\cite{Mueller2003}, the absence of data in a non-mandatory (nullable) attribute is often not considered an error in literature~\cite{Oliveira2005}. 
However, even if no constraints exist (e.g., in CSV or Excel files), missing values can still constitute errors when they represent real-world information that should be present.
}
% ----- Definition ------------------------------------------------------------
{\label{def:missing-val}
A value of an attribute $a$ for a tuple $t$ in a relation $r$ is considered erroneously missing if $ t.a.v = \bot \;\land\; M(t.a)\;\neq \bot$. That is, the value itself is the explicit null-value representation, but there exists a true value in the real-world.
}
% ----- Example ---------------------------------------------------------------
{
The tuple e1 in the \schema{Employee} relation has the \schema{HireDate} attribute set to $\bot$, resulting in an explicit missing value. 
As an employee, one should have a hire date, as reflected in the entity's real-world data in M(e1).
% , and it should be represented in their respective tuple in the relation.
}

\error[err:disguised-missing-val]
{Disguised missing value (DMV)}
% ----- Description ------------------------------------------------------------
{
A DMV represents missing data that is not explicitly recognized as a missing value for the respective attribute. Usually, DMVs can be easily recognized by humans because they follow identifiable patterns, deviate from the distribution of an attribute, or serve as well-known placeholders for missing values (e.g., ``\data{Unknown}'' and $-99$).
However, their detection can be challenging for automated approaches because they often fall within the defined domain or expected context of the attribute, thus appearing valid. Their detection can become extremely difficult or even impossible if they occur only once or a few times within an attribute and align with the overall distribution of the data. 
Their varied representation, the absence of global standards for their identification, and sometimes poor-quality metadata, such as missing, incomplete, or incorrect information about the attribute, significantly increase the complexity of detecting DMVs, resulting in additional effort during data cleaning~\cite{Pearson2006, Qahtan2018}. 
DMVs arise from different causes, such as mandatory attributes for which no real-world value exists at the time of data collection, users that enter fake values for privacy reasons, or users that insert default or guessed values. In this work, we define a DMV as a value that falls within the  domain and appears syntactically valid, but whose true real-world counterpart is a missing value, i.e., $M(\bot)$ as defined in \autoref{subsection:Preliminaries}, rather than a concrete real-world value. Values that fall outside the domain are treated as domain constraint violations (cf.\ \autoref{def:domain-const-violation}).
}
% ----- Definition ------------------------------------------------------------
{\label{def:disguised-missing-val} 
A value of an attribute $a$ for a tuple $t$ in a relation $r$ is considered a disguised missing value if $t.a.v = M(\bot) \;\land\;  t.a.v \in dom(a)$.

% \textcolor{blue}{Option 2}
% A value of an attribute $a$ for a tuple $t$ in a relation $r$ is considered a disguised missing value if $M(r.t.a) = \bot\;\land\; r.t.a \in dom(a)$.

% \textcolor{blue}{Option 3}
% A value of an attribute $a$ for a tuple $t$ in a relation $r$ is considered a disguised missing value if $M(r.t.a) = \bot\;\land\; r.t.a \in dom(a) \;\land\; r.t.a \neq \bot$.
}
% ----- Example ---------------------------------------------------------------
{
The tuple e2 in \schema{Employee} relation had \schema{TaxID} attribute value \data{``12 345 678 901''},  which is a syntactically correct but semantically incorrect value, as the hired person had not provided a tax ID at the time of hiring. 
The placeholder value satisfies the \schema{NOT NULL} constraint but replaces an explicit missing value with a disguised missing value.}

\error[err:partial-empty-tuple]
{Partial-empty tuple/attribute}
% ----- Description ------------------------------------------------------------
{
A tuple or an attribute in a relation is considered partially empty if the total number of cells with missing values (either \nullval or a disguised missing value) exceeds a predefined threshold~\cite{Oliveira2005}. 
The threshold is typically a fixed number or a percentage, defined depending on a particular use case. It can be the same for both tuples and attributes, or determined individually, depending on the context or specific DQ requirements.
}
% ----- Definition ------------------------------------------------------------
{\label{def:partial-empty-tuple} 

Let $\theta_t$ be a predefined threshold. A tuple $t$ is considered partially empty if $Count_{MV}(t.v) > \theta_t$.
Similarly, let $\theta_a$ be a predefined threshold. 
An attribute $a$ is considered partially empty if $Count_{MV}(a.v) > \theta_a$.
}
% ----- Example ---------------------------------------------------------------
{
The following are examples of partially empty tuples and attributes:

\begin{itemize}
  \item Let  $\theta_t = 0.25$ be the tuple-threshold on \schema{EmployeeCertificate} relation, which has four attributes. Then, the ec2 tuple is considered partially empty because it contains more than one missing value. 
  %is partially empty because two values are \NULL.
  \item Similarly, let $\theta_a = 0.3$ be the attribute threshold for the \schema{ManagerID} attribute in the \schema{Department} relation. The \schema{ManagerID} attribute is considered partially empty, as of the seven values in the attribute, three (for tuples d2, d4, and d7) are missing.
\end{itemize}
}

\error[err:missing-tuple]
{Missing tuple}
% ----- Description ------------------------------------------------------------
{
A tuple is considered missing when it is expected to be present (the corresponding real-world entity exists), but is absent from the relation~\cite{Oliveira2005, Mueller2003}. 
The absence of the expected tuples indicates an incomplete representation of real-world entities in a relation.
These missing tuples typically cannot be detected directly by examining a relation and require comparison with existing data using external sources or predefined expectations~\cite{Mueller2003}. 
Recall the functions $M(r)$, which maps relation $r$ to its real-world counterpart, as well as $M^*(r)$, which returns all expected real-world counterparts.
So, if $M(r) = M^*(r)$, then the relation $r$ has all the tuples that it is supposed to represent, and it is complete with respect to its intended population.
}
% ----- Definition ------------------------------------------------------------
{\label{def:missing-tuple}
A relation $r$ has missing tuples if and only if $|M^*(r) \setminus M(r)| > 0$.
}
% ----- Example ---------------------------------------------------------------
{
Our company acquires a start-up and integrates information about the new employees into the existing \schema{Employee} relation. 
Unfortunately, the start-up did not record training data; therefore, no related tuple exists for the relation \schema{Certificate}. 
This results in reduced year-end bonus payments for newly hired employees, regardless of their actual skills.
}

\error[err:empty-attribute]
{Empty attribute}
% ----- Description ------------------------------------------------------------
{
From a data profiling perspective, Abedjan et al.~\cite{Abedjan2015} describe attribute incompleteness as the presence of explicit missing values, where cells containing \NULL or empty strings are considered to have missing values. 
This captures situations in which only some, or possibly all, values of an attribute are missing.
For example, when tuples are inserted into a relation through views that do not include all attributes in their query, this results in explicit missing values for the omitted attributes~\cite{Grahne2018}.
From an attribute-level perspective, we extend this notion by considering the case in which all values in an attribute are missing either explicitly as \NULL or implicitly as \glspl{DMV}.
A common cause of empty attributes is poor schema design, where too many attributes are defined in a relation but are rarely populated.
}
% ----- Definition ------------------------------------------------------------
{
\label{def:empty-attribute} An attribute $a$ in a relation $r$ with $n$ tuples is considered empty if $Count_{MV}(a.v) = n$.
}
% ----- Example ---------------------------------------------------------------
{
Consider a view \schema{EmployeeView} defined over \schema{Employee}. A new attribute \schema{Phone} is added to the \schema{Employee} relation, but the view definition is not updated to include this attribute. As a result, \schema{Employee} relation shows \data{NULL} values for all phone numbers. Although valid phone number values exist for all employees in the real world, they are not propagated to \schema{EmployeeView} due to the outdated view definition.
}

% ----- Biased Data -----------------------------------------------------
\error[err:biased-data]
{Biased data}
% ----- Description ------------------------------------------------------------
{
The data in a relation have a natural underlying distribution when representing an external population or process. 
When tuples corresponding to certain classes or groups are missing or under-represented, the observed distribution becomes skewed, leading to biased representations. 
Such missing data introduces information gaps by removing evidence about parts of the population and shifting the observed distribution, which can lead to discriminatory outcomes in downstream tasks for underrepresented classes or groups~\cite {Emilio2024}.
Bias can be caused by subjective judgment during data collection~\cite{Lee2006, Emilio2024} (also known as cognitive bias), selective data availability due to privacy and security constraints~\cite{Kurapati2025}, intentional exclusion of certain groups during data entry~\cite{Kurapati2025} (also known as exclusion bias), as well as missing data mechanisms such as MAR and MNAR. Note that MCAR does not necessarily introduce bias. 
If the real world itself has uneven group sizes, even complete data representation can under-represent certain classes. 
For example, breast cancer datasets naturally contain many fewer male than female patients, not because of missing tuples but due to actual disease distribution in the real world. 
While this imbalance may not constitute a data error, given that the data are complete, it should be addressed based on the intended task. Therefore, we treat bias as an error indicator, as it signals potential issues in the data but requires context and judgment to determine whether it constitutes an error in the data.
To formally define bias on the database element $e$, we use two distribution functions, $P_{obs}(e)$, which returns the observed distribution and $P_{exp}(e)$, which returns the expected distribution of element $e$. 
To measure bias in data, we also define a distance function $D(P_{obs}(e), P_{exp}(e))$, where $D$ can be calculated using common distance measures such as Kullback–Leibler divergence, Jensen–Shannon divergence, total variation distance, or Wasserstein distance.
}
% ----- Definition ------------------------------------------------------------
{
A set of attributes $A_s \subseteq A$ is considered to have bias if $D(P_{obs}(A_s), P_{exp}(A_s)) > \theta$, where $\theta$ is predefined bias threshold.
}
% ----- Example ---------------------------------------------------------------
{
The \schema{EmployeeCertificate} relation is intended to record all completed certification data for both permanent and contract-based employees. However, due to inconsistent reporting practices, certifications for many contract-based employees across multiple departments were not recorded. 
These missing tuples introduce bias, leading to a substantial deviation from the expected distribution across employment types. 
}
\section{Incorrect Data}
\label{sec:incorrectdata}
\glsreset{RDB}
\glsreset{OOV}
\glsreset{FD}

Data in a relational database (RDB) are considered incorrect when they do not accurately represent the intended real-world entity or its relationships. 
Based on semantic validity, a data value can either be invalid (outside the predefined domain or logically incorrect in the real world), e.g., value \data{-54000} for an employee salary, or valid (within the expected domain) but still incorrect, e.g., value \data{10000} entered as salary where it is in fact \data{54000}.
For better readability, we further group incorrect data according to their nature into four families as shown in the ~\autoref{tab:errorclassification}:
\begin{enumerate}
    \item \emph{Textual errors}, also known as spelling errors, 
    %are specific to \data{string} data types 
    with examples like \gls{OOV} words, misspellings, or typos.
    \item \emph{Nonconformant errors} describe outlying or clearly anomalous data, including contradictions, formatting issues, and outliers.
    \item \emph{Rule violations} are logic-based error indicators that capture violations in specific rules or constraints, such as a business rule or a functional dependency violation.
    \item \emph{Outdated data} is considered to be present in a relation when it is not sufficiently up to date with the current or any previously valid schema and therefore not fit for any task at hand.
\end{enumerate}

% ----- INVALID VALUE/TUPLE -----
\error[err:invalid-val-tup]
{Invalid value and tuple}
% ----- Description ------------------------------------------------------------
{
A data value is considered invalid if it violates schema constraints (e.g., type, domain, or format) or is illogical in that attribute's context (e.g., negative salary). 
We discuss various schema constraint violations in their respective sections. 
In this section, we focus on invalid values and tuples that result from deviations from the expected context of an attribute or tuple. 
The term \et{invalid value} is often used as an umbrella term for several error types, and its presence is often a sign of broader data quality issues.
Oliveira et al.~\cite{Oliveira2005}, for instance, discuss ``invalid substrings'', which occur when a data value falls outside the intended semantics of an attribute, even though part of it still adheres to those semantics, e.g., some \schema{FullName} values in \schema{Employee} relation include titles like ``\data{Prof.}'' or ``\data{Dr.}''. 
In contrast, Curran~\cite{Curran2016} and Johnson~\cite{Johnson2005} study invalid values from a behavioral perspective and describe them as data that fail to represent a respondent’s actual values, often due to misunderstanding or carelessness during the data collection process.
While this perspective captures a broader notion of semantic inaccuracy, it extends beyond how invalidity is typically defined in data management. 
The focus usually lies on explicit violations of schema or logical constraints.
In this work, we adopt the data management perspective, aligning more closely with Oliveira et al.~\cite{Oliveira2005}.

Invalid tuples extend the notion of invalid values to entire tuples in a relation. Jensen and B{\"{o}}hlen~\cite{Jensen2002} consider a tuple invalid if ``it is neither a current nor a legacy tuple'', that is, when it no longer conforms to the current or any previous valid schema version. 
Müller and Freytag~\cite{Mueller2003} take a semantic perspective and consider a tuple invalid if it does not represent a valid entity in the modeled reality. 
Josko et al.~\cite{Josko2016} refer to invalid tuples as false tuples, which include tuples that do not represent valid real-world entities in the modeled domain.
Such cases are difficult to detect, as they might not violate any explicit integrity constraints.
In practice, invalid tuples may arise from a variety of underlying issues, such as data type mismatches, inconsistent formatting, invalid attribute values, or spelling errors. 
Unlike explicit constraint violations, invalid values and tuples are often not automatically detected by the database management system (DBMS), since they may also fall outside the scope of predefined constraints. 
% Their identification typically requires additional validation, domain knowledge, or custom rules defined for the attributes. 
}
% ----- Definition ------------------------------------------------------------
{
\label{def:invalid}
Let $\textit{Valid}(e)$ be a Boolean function that evaluates whether a database element $e$ satisfies all relevant validity conditions. Then a data value $t.a.v$ is considered invalid if $t.a.v \neq \bot\;\land\; \neg Valid(t.a.v)$.
Similarly, a tuple in relation $r$ is considered invalid if $\neg \textit{Valid}(r.t.v)$.
}
{
The following are examples of an invalid value and a tuple from the \schema{Employee} relation:

\begin{itemize}
  \item The tuple e3 in \schema{Employee} relation has a salary of \data{-65,000}, which is invalid, since salaries cannot be negative in any real-world context.
  \item The tuple e4 in \schema{Employee} relation is considered invalid due to multiple invalid values, including a numeric character in the full name, improper formatting of the email address, and a textual value in the salary field, which violates the current as well as any legacy employee schema.
\end{itemize}
}

% ----- OOV WORDS -----
\error[err:OOV-word]
{Out-of-vocabulary (OOV) word}
% ----- Description ------------------------------------------------------------
{
A word is considered (OOV) if it is not found in the vocabulary of the selected language~\cite{Chen2019}.
The term vocabulary is defined as ``all the words in a particular language'' or ``all the words that a person knows or uses''~\cite{NOAD}. 
Such a vocabulary can be a traditional dictionary, a predefined set of words for a specific domain, or simply a formal grammar that defines valid words. 
A vocabulary is the explicit enumeration of all  allowed words (as opposed to a domain specification cf.~\autoref{def:domain-const-violation}). 
We use the $OOV(w, V_a)$ function (defined in \autoref{subsection:Preliminaries}) for a formal definition, which, if $\true$, indicates that the word is not found in the selected vocabulary. 
}
% ----- Definition ------------------------------------------------------------
{
\label{def:oov}
A value or part of a value, denoted by $w$, of an attribute $a$ for a tuple $t \in r$ is considered out-of-vocabulary (OOV) if $w \neq \bot \;\land OOV(w, V_a) = \true$, where $V_a$ denotes the vocabulary associated with attribute $a$. 
}
{
The tuple e2 of \schema{Employee} relation has value``\data{Mgr.}'' in \schema{Role} attribute.
As this abbreviation does not exist in the defined English vocabulary, ``\data{Mgr.}'' is considered an \gls{OOV} word.
}
% integrity (different kinds in order) - referential - fd - CFD

\error[err:spelling]
{Misspelling, typo, misscan, and incorrect encoding}
% ----- Description ------------------------------------------------------------
{
Errors in the categories of misspellings, typos, misscans, and incorrect encoding share a common property; they arise from unintended alterations at the character level of a value. 
These alterations can be caused by insertion, substitution, omission, or transposition in one or more characters of the correctly spelled word~\cite{Damerau1964}.  
The existing research is primarily focused on the detection and correction of these errors, specifically in string values~\cite{Damerau1964, Kukichq1993, Mazeika2006}. However, typos, misspellings, and incorrect encoding errors are not restricted to just strings.
For example, a decimal separator may be mistyped, resulting in a floating-point value that is syntactically valid but incorrect. 
In such cases, downstream systems often flag the issue as a formatting error because the underlying cause cannot be precisely identified. 
For this reason, and unlike prior work, we do not limit our discussion to string attributes for these three error types. 
Instead, we adopt a standard definition, treating them as manifestations of character-level deviations, regardless of their distinct, often unobservable causes. 
Since these causes are diverse and challenging to identify during data cleaning, they cannot be meaningfully included in a formal definition.

All spelling deviations across these four categories can be detected at two levels: in OOV words or non-OOV words. 
Intuitively, a word that does not appear in a standard dictionary indicates that it is not a valid term (e.g., the \schema{DeptName} value \data{Information Technology} misspelled as \data{Informasion Technology})~\cite{Oliveira2005}. 
This type of deviation is also referred to as a \emph{non-word} error.
However, just because a word is \gls{OOV}, this does not necessarily imply an error, since many valid names (e.g., \data{X Æ A-12}, which is the name of Elon Musk's child), codes, company- and domain-specific terms (e.g., \data{Fédération Cynologique Internationale} as \schema{DeptName}) are also outside standard dictionaries. 
To check whether an OOV word contains a textual error, we define a 
%To express such cases formally, we define a 
textual error type identification function, $TEI(w)$, inspired by Norvig’s probabilistic spelling corrector~\cite{Norvig2007}. 
%The $TEI$ function models the detection of character-level deviations using edit distance, word frequency, and statistical and \gls{ML} techniques. 
Hence, we use the $OOV$ (see \autoref{subsection:Preliminaries}) function together with the $TEI$ function for identifying non-word spelling errors.
However, even a correctly spelled word can be erroneous if it is affected by a spelling error, e.g., the employee role \data{Manager} written as the valid word \data{Manger}~\cite{Jurafsky2009}. 
These correctly spelled cases are also known as \emph{real word} spelling errors~\cite{Wilcox2008}. 
Such an erroneous value appears valid in isolation and is only semantically incorrect (i.e., used in the wrong context).
%Factors such as letter arrangement and common phonetic confusions contribute to the production of such errors~\cite{Kukichq1993, Yanna1983}. 
To express such cases formally, we define a context-based textual error identification function, $CTEI(w, e_{context})$, inspired by Mays et al.~\cite{Mays1991}, where $e_{context}$ (usually the whole tuple) represents the surrounding text used to assess contextual fit. 
A word that is valid in a selected vocabulary but contextually incorrect is flagged as a potential spelling error by the $CTEI$ function. 
In the following paragraphs, we describe each spelling error in more detail. 

\textit{Misspelling}: Misspellings result in data when a person does not know the correct spelling of a word or incorrectly recalls it during data entry. 
In many cases, the person relies on how the word sounds, which introduces cognitive confusion when words have similar pronunciations or when their native language shapes how they perceive sounds~\cite{Cox2017}. These errors are often not random. 
They tend to occur repeatedly in the same word because the person entering the data consistently spells them incorrectly. 
Misspellings often follow specific patterns, such as swapped letters, omitted letters, or phonetic substitutions, thus creating a major problem in large-scale text processing~\cite{Popescu2014}. 

\textit{Typos}: A typographical error or typo occurs when the spelling of a word becomes incorrect due to accidentally pressing the wrong key during typing. These errors often arise from fast typing or a finger slipping on the keyboard, causing a character to be replaced by a nearby key, repeated unintentionally, or omitted entirely from the word.
Typos commonly follow five basic forms of mistyping: substitution, insertion, deletion, replication, or transposition~\cite{Shah2020}. 

\textit{Misscan}: Misscan occurs when an Optical Character Recognition (OCR) software~\cite{Islam2017} incorrectly scans and interprets text from text files or images, resulting in an inaccurate distinction of different characters~\cite{Kukichq1993}. Such scanning errors differ from other spelling errors (misspellings, typos, and encoding errors) because they have different causes and varying patterns of occurrence, such as \data{O} being interpreted as the number \data{0}. Misscans can be divided into four categories based on the underlying issues within the OCR pipeline: (1)~failures in text detection, (2)~character recognition, (3)~word boundary detection, and (4)~character boundary detection~\cite{Kissos2016}.

\textit{Incorrect Encoding}: An incorrect encoding error occurs when data does not conform to the expected format due to an inaccurate or incompatible transformation from one encoding format to another~\cite{Kim2003}. 
Errors in this category arise from issues with how data are represented and transformed across formats, rather than from software misinterpreting characters or a person mistyping a value.
}
% ----- Definition -------------------------------------------------------------
{
\label{def:spelling-err} A value or part of a value, denoted by $w$, of an attribute $a$ for a tuple $t \in r$ is considered a spelling error if $(OOV(w, V_a) = \true \;\land\; TEI(w) = \true) \;\lor\; (OOV(w, V_a) = \false \;\land\; CTEI(w, t.v) = \true)$.
}
{The following are examples of each textual data error:
\begin{itemize}
  \item \textbf{Misspelling:} The full name of employee ``\data{Sara Müller}'' can be incorrectly recorded via manual data entry as ``\data{Sera Müller}'', due to the similar pronunciation of the names ``\data{Sara}'' and ``\data{Sera}'' in English. Although both are valid first names, this misspelling reflects a real-world error caused by phonetic confusion.
  \item \textbf{Typo:} The full name of employee ``\data{Sara Müller}'' can be mistyped as ``\data{Ssara Müller}'' where the first character ``s'' was accidentally typed twice during typing.
  \item \textbf{Misscan:} Some names in \schema{FullName} attribute containing letters ``ü'' got improperly scanned resulting in OCR errors like name ``\data{Sara Müller}'' in tuple e5 was incorrectly scanned as ``\data{Sara Moller}''.
  \item \textbf{Incorrect Encoding:} The full name of an employee ``\data{Sara Müller}'' can appear as ``\data{Sara M\&\#$252$; ller}'' if the employee's digital documents were converted to HTML format before they were entered into \schema{Employee} schema.
\end{itemize}}

% ----- SYNONYMS ---------------------------------------------------------
\error[err:synonyms]
{Synonym}
% ----- Description ------------------------------------------------------------
{
Synonyms are lexical relationships in which different words or textual representations refer to the same entity or concept. These include vernacular words, acronyms, abbreviations, symbols, or standard words in a language that point to the same concept~\cite{Josko2016}. 
These can also be referred to as lexical variations of a spelling, which can result in semantic ambiguity in the data. 
Synonyms occur for various reasons, such as data integration from multiple sources that use different lexical forms for the same value, or through inconsistencies introduced by manual entry. They are also common in short texts on social media, where users often prefer abbreviations or alternative terms to fit within character limits~\cite{Mu2023}.
The presence of synonyms may also be a cause for duplicate tuples (cf. ~\autoref{err:duplicate-tuples}). For example, employee records from different companies may use synonymous country values such as GER, DE, and Germany. Since these values cannot be automatically matched, they result in duplicate tuples referring to the same country.
}
% ----- Definition ------------------------------------------------------------
{
\label{def:synonyms}Let the symbol $\approx_s$ denote semantic similarity, indicating that two words that refer to the same entity or concept but have different lexical representations. Let $w$ be a value or part of a value in an attribute $a$ for tuple $t$, where $D_{type}(a) = string$.
The word $w$ is considered a synonym in place of the expected word when $CTEI(w, t.v) = \true \;\land\; w \approx_s M(t.a)$.
}
% ----- Example ---------------------------------------------------------------
{
The \schema{Department} relation has tuples d1, d5, and d6, which refer to the same department \data{Human resource}, but represented with semantically synonymous names. Employees in the \schema{Employee} relation may be associated with any department name, leading to inconsistencies due to synonym usage.}

% ----- WORD TRANSPOSITION -----
\error[err:word-transposition]
{Word transposition}
% ----- Description ------------------------------------------------------------
{
A word transposition occurs when values within a multi-value or free-form attribute (such as \schema{FullName}) are interchanged, with each being entered into the position of the other~\cite{Kim2003, RahmDo2000}. These transposed values are individually valid, but in an attribute that expects them to be in a particular order, they have been placed in the wrong positions. Note that misplacement into wrong attributes is a \et{misfielded value} and explained separately in \autoref{err:misfielded-val}.
Such errors are often observed in manually entered data and can hinder the correct interpretation of the information and cause inconsistencies across the system.
}
% ----- Definition ------------------------------------------------------------
{
\label{def:word-transposition}Let $[w_1, w_2,\ldots, w_i,\ldots, w_n]$ denote the expected order of words in a free-form attribute $a$ using positional index $i$ $(0 \leq i \leq n)$. 
If any word deviates from this expected order, the resulting sequence of words in $a$ for a tuple $t$ is considered transposed.
}
% ----- Example ---------------------------------------------------------------
{
The attribute \schema{FullName} in tuple e6 of the \schema{Employee} relation has first and last names transposed, resulting in full names as ``\data{Bond James}'' instead of ``\data{James Bond}''. 
}

% ----- MISFIELDED VALUES -----
\error[err:misfielded-val]
{Misfielded value}
% ----- Description ------------------------------------------------------------
{
A misfielded value occurs when a semantically valid value is recorded under the wrong attribute of a relation~\cite{RahmDo2000}. 
%, e.g., \schema{Country} column has value \data{Berlin}
Such errors can be caused by data entry or software issues, where a value is incorrectly entered into an attribute to which it doesn't belong, and are challenging to detect if the values are misfielded between similar attributes, such as department ID (attribute \schema{DID} and manager ID (attribute \schema{ManagerID}) in \schema{Department} relation.
}
% ----- Definition ------------------------------------------------------------
{
\label{def:misfielded-val} Let $a_1$ and $a_2$ be two attributes in a relation $r$.
A misfielded value error occurs when a value belonging to $a_1$ is recorded under $a_2$, that is, $t.a_2.v \neq M(t.a_2) \;\land\; t.a_2.v = M(r.t.a_1)$.
}
% ----- Example ---------------------------------------------------------------
{
Some part of the \schema{FullName} of the employee in tuple e7 in \schema{Employee} relation was entered into \schema{HireDate} attribute due to improper handling of the apostrophe, resulting in a misfielded value error. 
}

% ----- NOISE -----
\error[err:noise]
{Noise}
% ----- Description ------------------------------------------------------------
{
Noise, in general, results from interference during data collection, causing the recorded values to deviate from the actual value. It can result from multiple factors, such as sensor failure, improperly calibrated measurement equipment, errors during data entry, and software issues~\cite{Subramaniam2009}. Such a broad definition of the term ``noise'' sometimes encompasses many other categories of data quality problems, including misspellings, formatting issues, miscounts, and outliers. We discuss these specific error types in their dedicated sections. Specifically, in signal processing, noise can be described as additive noise, multiplicative noise~\cite{Winkler2003}, or quantization error~\cite{Pesce2023}. When data from measurement systems is stored in tabular form, these formulations appear as unintended deviations in the recorded values.
}
% ----- Definition ------------------------------------------------------------
{
\label{def:noise}Let $\eta_{a,t}$ be a noise term representing any unintended deviation in the value of an attribute $a$ for tuple $t$ and let $\oplus$ represent any operation through which this deviation is introduced. Then the attribute $a$ is said to contain noise if there exists a non-empty subset $t^{'} \subseteq r$ such that for each tuple $t \in t^{'}: t.a.v = M(t.a) \oplus \eta_{a,t}$.
}
% ----- Example ---------------------------------------------------------------
{
The system computes the values in the \schema{Salary} attribute of the \schema{Employee} relation based on the daily working hours captured by electronic time-tracking devices for contract-based employees. Minor timing inconsistencies in these devices, which remain unnoticed during regular operation, introduce random variation in the recorded hours. As a result, the salaries of the contract-based employees represented in the Salary attribute may fluctuate by several tens of dollars between payroll runs, even when the employees’ working patterns remain unchanged.
}

\error[err:semantic-amb]
{Semantically ambiguous data}
% ----- Description ------------------------------------------------------------
{
Semantically ambiguous data occurs when the value of an attribute or the values in a tuple are not sufficient to uniquely identify the underlying real-world entity. Such ambiguity can appear in a relation in two forms. 
At the attribute level, this manifests as \et{imprecise values}~\cite{Oliveira2005}, such as abbreviations, shorthand notations, or homonyms~\cite{Oliveira2005}, that could refer to multiple valid domain values. While abbreviations and shorthand notations can also appear as lexical variation errors (e.g., synonyms or alternative representations of the same concept), here we consider cases where they correspond to multiple possible meanings.
For example, the value \et{DMV} may correspond to disguised missing values or the Department of Motor Vehicles. 
At the tuple level, the representation of an entity is ambiguous~\cite{Talburt2013} if a single record could plausibly refer to multiple distinct real-world entities. 
In both cases, the mapping between the information system and the real world becomes one-to-many rather than one-to-one. 
Such data prevents accurate real-world entity identification and can result in inconsistencies in downstream analysis.
}
% ----- Definition ------------------------------------------------------------
{
\label{def:semantic-amb}A value in an attribute $a$ or a complete tuple $t$ is considered semantically ambiguous if it maps to more than one real-world element or fact, that is, $|M(t.a)| > 1 \;\lor\; |M(t)| > 1$.
}
% ----- Example ---------------------------------------------------------------
{
The attribute \schema{CertName} of tuple c1 in \schema{Certificate} relation contains the value \data{SAP HANA Certification}. The name of the certificate creates semantic ambiguity because SAP offers multiple certifications with nearly identical titles, such as SAP HANA 1.0, SAP HANA 2.0, SAP HANA 2.0 SPS05, or edition-specific credentials. Since the value \data{SAP HANA Certification} can plausibly refer to multiple real-world certificates, the tuple cannot be mapped uniquely to a single credential.
}

% ----- OUTLIER -----
\error[err:outlier]
{Outlier}
% ----- Description ------------------------------------------------------------
{
Hawkins et al.~\cite{Hawkins2002} define an outlier as ``an observation that deviates so much from other observations as to arouse suspicion that a different mechanism generated it''. 
In general, a value is considered an outlier if it significantly differs from the distribution formed by surrounding data, and this distribution can be defined by values in a single attribute (cf. univariate outlier) or multiple attributes (cf. multivariate outlier) combined together~\cite{Sullivan2021, Ilyas2022}. 
Such atypical values appear under various names in literature, including anomalies, exceptions, deviants, and aberrations; however, outliers and anomalies are the most commonly used terms~\cite{Chandola2009}. 
Outliers differ from noise. Noisy values may lie close to the primary distribution of the data, yet still be inaccurate due to small measurement or recording errors. Outliers, in contrast, may be valid but rare observations, so their presence does not necessarily indicate an error and often demands careful interpretation~\cite{Salgado2016}.
}
% ----- Definition ------------------------------------------------------------
{
\label{def:outlier}
Outliers are data values that significantly deviate from the overall distribution in one or more attributes.
}
% ----- Example ---------------------------------------------------------------
{
The \schema{Salary} attribute of the employee in tuple e7 has value \data{205,000}, whereas the value of other employees typically ranges between \data{40,000} and \data{80,000}. The substantial deviation of this employee's salary from the majority of values in the \schema{Salary} attribute indicates an outlier and requires further investigation to determine whether it is an error.
}

% ----- SYNTAX VIOLATION -----
\error[err:syntax-viol]
{Syntax violation}
% ----- Description ------------------------------------------------------------
{
A syntax violation occurs when values in an attribute do not conform to a predefined or expected syntax or grammar, and therefore it cannot be parsed by a syntax parser~\cite{Oliveira2005, Christian2011}. We use $Syntax(t.a.v)$ function described in \autoref{sec:preliminaries} for defining syntax violations.
}
% ----- Definition ------------------------------------------------------------
{
\label{def:syntax-viol}Let $S_a$ be the predefined syntax for an attribute $a$. A syntax violation occurs when the value of the attribute $a$ for a tuple $t$  does not conform to the syntax defined for the attribute $a$, that is, $Syntax(t.a.v) \neq S_a$.
}
% ----- Example ---------------------------------------------------------------
{
The \schema{HireDate} attribute has an expected syntax of \schema{dd/mm/yyyy}, but the hire date in tuple e4 in \schema{Employee} relation has the value with \schema{dd.mm.yyyy} syntax, which results in a syntax violation.
}

% ----- HETEROGENEOUS FORMATTING -----
\error[err:heterogeneous-formatting]
{Heterogeneous formatting}
% ----- Description ------------------------------------------------------------
{
A heterogeneous formatting error, also known as heterogeneity of syntax~\cite {Oliveira2005}, occurs when two attributes representing semantically similar data in one or more relations follow different syntax, even though they are meant to store the same data. 
These similar attributes may be found in a single relation, or in separate relations where the attributes representing similar data have different formats. Heterogeneous formatting errors differ from syntax violations, as the latter occur within the values of an attribute. In contrast, the former arise when attributes of the same or similar data types (e.g., dates) adopt inconsistent syntactic representations across the database.
}
% ----- Definition ------------------------------------------------------------
{
\label{def:heterogeneous-formatting}Let $a_1$ and $a_2$ be two attributes with syntax $S_{a_1}$ and $S_{a_2}$, respectively, that share the same semantic meaning, i.e., $a_1 \equiv_{s} a_2$ and are expected to have the same syntax. Heterogeneous formatting occurs when, $a_1$ and $a_2$ have different syntax, that is, $a_1 \equiv_{s} a_2 \;\land\; S_{a_1} \neq S_{a_2}$.

}
% ----- Example ---------------------------------------------------------------
{
The \schema{Employee} relation stores values in attribute \schema{HireDate} in \schema{dd/mm/yyyy} format, while attributes \schema{ExptCompDate} and \schema{ActCompDate} in \schema{EmployeeCertificate} relation are recorded in \data{mm/dd/yyyy} format, creating a heterogeneous formatting issue for the same data type within a database.
}

% ----- INCORRECT UNIT
\error[err:incorrect-unit]
{Incorrect unit}
% ----- Description ------------------------------------------------------------
{
An incorrect unit error occurs when the values of an attribute violate the unit semantics specified for that attribute. 
Typically, a schema defines the unit for an attribute, but individual records can still use different units to encode their measurements.
This category, therefore, also includes errors arising from values recorded at different abstraction levels, i.e., heterogeneous granularity~\cite{Josko2016}.
A mismatch in data recorded with a different unit than the expected unit can propagate through analytical or downstream systems, resulting in biased results. 
% Note that the incorrect unit error differs from a heterogeneous unit error. The latter arises when attributes with the same semantic roles across different relations in a database use different measurement units~\cite{Josko2016, Oliveira2005}, which is more of a metadata issue. 
We use the $Unit(a)$ function described in \autoref{subsection:Preliminaries} for defining an incorrect unit.

}
% ----- Definition ------------------------------------------------------------
{
%Let $a$ be a numeric attribute in relation $r$.
\label{def:incorrect-unit}An incorrect unit error occurs when $\textit{Unit}(a) \neq M(\textit{Unit}(a))$.
}
% ----- Example ---------------------------------------------------------------
{
The salary values for some employees in the \schema{Employee} relation are recorded in dollars instead of the expected unit of euros. 
As a result, the salary value is numerically plausible but violates the attribute's unit semantics.
}

% ----- INCORRECT REFERENCE -----
\error[err:incorrect-ref]
{Incorrect reference}
% ----- Description ------------------------------------------------------------
{
An incorrect reference occurs when a foreign key~\cite{Ordonez2008} in one relation points to an existing primary key in another relation, but the referenced tuple does not correspond to the correct real-world entity intended by the referencing tuple~\cite{Oliveira2005, Josko2016}. 
An incorrect reference differs from a referential integrity violation as the former is syntactically valid but semantically incorrect, whereas the latter occurs when the referenced attribute does not exist in the corresponding relation. 
}
% ----- Definition ------------------------------------------------------------
{
% \label{def:incorrect-ref}Let $r_1$ and $r_2$ be two relations, such that $r_2$ contains a non-empty set of foreign key attributes $A_{r_2,FK}$, referencing the primary key attributes $A_{r_1,PK}$ in $r_1$. An incorrect reference occurs when the values for foreign key attributes in tuple exist in the referenced primary key but does not correspond to the correct real-world entity, that is, $r_2.t.A_{r_2,FK} \in r_1.A_{r_1,PK} \;\land\; r_2.t.A_{r_2,FK} \neq M(r_2.t.a_{r_2, FK})$.

\label{def:incorrect-ref}
Let $r_1$ and $r_2$ be two relations, where a foreign key (can also be composite key) $A_{FK}$ in $r_2$ references the primary key $A_{PK}$ in $r_1$. An incorrect reference occurs for a tuple $t \in r_2$ when the foreign key value exists in the referenced relation but does not correspond to the correct real-world entity, that is,
$
r_2.t.A_{FK}.v \in r_1.A_{PK}.v \;\land\; t.A_{FK}.v \neq M(t.A_{FK}).
$
}
% ----- Example ---------------------------------------------------------------
{
The \schema{DID} attribute for the employee in tuple e7 in the \schema{Employee} relation references values from \schema{DID} attribute in \schema{Department} relation. This employee actually belongs to department \data{30}, but the tuple contains a value \data{10}, resulting in an incorrect reference to the \schema{DID} value.
}

% ----- CONSTRAINT VIOLATION -----
\error[err:const-violation]
{Constraint violation}
% ----- Description ------------------------------------------------------------
{
Constraints are formal rules that ensure that the database is in its correct semantic state (consistent with the meaning intended by its schema)~\cite{Grefen1993}. A constraint violation occurs when the data in a relation violates any of these defined formal rules~\cite{Decker2011}. These formal rules serve as safeguards to ensure that authorized operations, such as insertions, deletions, or updates, do not compromise the database's consistency. Hence, a constraint violation occurs when an operation creates a database state that fails to satisfy the semantics prescribed by the schema. From a broader perspective, consistency also refers to the validity of data with respect to the real world it represents, as emphasized by Fan~\cite{Fan2015}, who views constraint violations as indicators of inconsistencies in that representation.
Constraint violations are also popularly referred to as \et{integrity constraint violations}~\cite{Mueller2003} or \et{data dependency violations}~\cite{Fan2008DC}.

The various types of integrity constraints~\cite{Grefen1993, Hector2009} include classical forms, such as domain, \schema{NOT NULL}, uniqueness, and referential integrity constraints, as well as extended forms, including recursive, transition, temporal, and fuzzy constraints.
In addition, relationship-level constraints such as inclusion, cardinality, and participation constraints have also been studied~\cite{Josko2016}.
In this section, we provide a unified definition of all integrity constraint violations, with dedicated subsections on uniqueness and domain constraint violations. 
We refer the interested reader to \cite{Fan2008DC, Decker2011, Fan2012} for details on integrity constraints.
}
% ----- Definition ------------------------------------------------------------
{
\label{def:const-violation}
Let $ConstraintCheck(e)$ be a boolean function that returns \true\ if all constraints that apply to the database element $e$ are satisfied. A constraint violation at database element level $e$ occurs if $\textit{ConstraintCheck}(e) = \false$.
}
% ----- Example ---------------------------------------------------------------
{
In the \schema{Department} relation, an integrity constraint requires that each department have only one manager.
In this relation, tuples d1 and d6 violate the integrity constraint, since the same department (\schema{DID} = 20) is associated with two different managers (\schema{ManagerID} = 103 and 105).
}

% ----- DOMAIN CONSTRAINT VIOLATION
\error[err:domain-const-violation]
{Domain constraint violation}
% ----- Description ------------------------------------------------------------
{
A domain is a model of all allowed values for an attribute, e.g., \schema{int}, \schema{string}, or the range specified for an attribute in the database schema~\cite{Oliveira2005}.
A domain constraint violation is the most basic form of integrity constraint violations and occurs when the data does not conform to the model.
% set of valid values (i.e., set violation~\cite{Oliveira2005b}) or the 
Note that domain constraints differ from data type defined on an attribute, as a domain may restrict values beyond the data type. 
If undefined, the domain of an attribute includes all valid values for its data type~\cite{Grefen1993}. If explicitly defined, the domain represents a restricted subset of the data type, typically specified by constraints.
}
% ----- Definition ------------------------------------------------------------
{
\label{def:domain-const-violation}
Let $dom(a)$ denote the domain of an attribute $a$ in relation $r$, which specifies the set of valid atomic values of $a$. 
A domain constraint violation occurs when a tuple $t$ in the relation $r$ contains a value for the attribute $a$ such that $t.a.v \notin dom(a)$.
}
% ----- Example ---------------------------------------------------------------
{
The domain of the \schema{DeptName} attribute in the \schema{Department} relation is defined as the set of values, \{\data{`Human Resource', `IT', `Management', `Security'}\}. 
Tuple d7 has a department name that falls outside the attribute's domain, resulting in a domain constraint violation.
}

% ----- UNIQUENESS VIOLATION -----
\error[err:uniqueness-violation]
{Uniqueness violation}
% ----- Description ------------------------------------------------------------
{
A uniqueness violation occurs when two or more tuples in a relation share the same value for an attribute (or set of attributes) that is required to be unique, and is declared as unique using the \schema{UNIQUE} constraint~\cite{Oliveira2005, Christian2011}. The attributes with \schema{UNIQUE} constraint do not necessarily form a key, as they are not required to uniquely identify tuples or be non-null. They only enforce that duplicate values are not allowed.
}
% ----- Definition ------------------------------------------------------------
{
\label{def:uniqueness violation} Let $A_{uniq} \subseteq A$ be a set of nonempty attributes in a relation $r$ for which a \schema{UNIQUE} constraint is defined, or unique values are expected.
A uniqueness violation occurs when two tuples $t_1, t_2 \in r$ have the same value combination for the attribute set $A_{uniq}$, that is, $t_1.A_{uniq}.v = t_2.A_{uniq}.v$.
}
% ----- Example ---------------------------------------------------------------
{
The \schema{Email} attribute in the \schema{Employee} relation has a \schema{UNIQUE} constraint; however, two tuples, e1 and e9, share the same \schema{Email}, resulting in a uniqueness violation.
}

% ----- ATTRIBUTE DEPENDENCY VIOLATION -----
\error[err:attr-dep-violation]
{Attribute dependency violation}
% ----- Description ------------------------------------------------------------
{
An attribute dependency violation occurs when the data does not adhere to the predefined relationships or dependencies between two or more attributes in a tuple. 
Rahm and Do~\cite{RahmDo2000} refer to such cases as violated attribute dependencies, which may arise at both single-source and multisource levels. Common subcategories include violations of functional dependencies, conditional functional dependencies~\cite{Oliveira2005, Hector2009, Papenbrock2015, Josko2016}, join dependencies (JDs)~\cite{Abiteboul1995, Kolahi2009}, multivalued dependencies (MVDs)~\cite{Abiteboul1995, Kolahi2009}, inclusion dependencies (INDs)~\cite{Fan2012, Josko2016}, and conditional inclusion dependencies~\cite{Fan2012}.
We discuss functional dependency (FD) violations in \autoref{err:fd-violation} and conditional FD violations in \autoref{err:cfd-violation} as they are widely studied and have well-established definitions, and give a general definition for attribute dependency violation using the attribute dependency check function defined in \autoref{subsection:Preliminaries}.
}
% ----- Definition ------------------------------------------------------------
{
\label{def:attr-dep-violation}
Let $A_{lhs}, A_{rhs} \subseteq A$ be non-empty sets of attributes in a relation $r$, where an attribute dependency is defined as $A_{lhs} \xrightarrow{c} A_{rhs}$ where $c$ denotes an optional condition. An attribute dependency violation occurs if $DependencyCheck(A_{lhs}, A_{rhs}, c) = \false$.
}
% ----- Example ---------------------------------------------------------------
{
In the \schema{Employee} relation, the value of the \schema{Role} attribute is semantically dependent on the \schema{DID} attribute. In tuple e7, the \schema{DID} value is \data{10}, while the corresponding \schema{Role} value is ``\data{Head of Security}'', which is inconsistent with the expected department-role mapping. This constitutes an attribute dependency violation.
}

% ----- FUNCTIONAL DEPENDENCY VIOLATION -----
\error[err:fd-violation]
{Functional dependency (FD) violation}
% ----- Description ------------------------------------------------------------
{
FDs are constraints in which a set of non-empty attributes determines the values of another set of attributes in a relation~\cite{Ramakrishnan2003, Oliveira2005, Hector2009, Papenbrock2015, Josko2016}. 
They capture semantic relationships between attributes and play a central role in schema normalization and data integration~\cite{Thorsten2016}.
Although FDs can involve sets of attributes on either side, in the simplest case, an FD can be defined between two attributes: if two tuples agree on the value of one attribute, then they must also agree on the value of the second attribute~\cite{Golab2008}. 
When data does not follow these constraints, FD violations occur. 
Such violations are frequently observed when integrating data from multiple sources, extracting data from the Web, or working with complex and evolving schemas that make it difficult to define all possible FDs~\cite{Beskales2010}. 
Note that trivial FDs do not constitute any dependency violations~\cite{Golab2008}.  
Conditional FD violations generalize FD violations, whereas disjoint subdomains, incompatible replication, and inference rule violations can be seen as semantic or structural variants of FD-like consistency violations~\cite{Josko2016}.

% Also, multivalued dependencies \glspl{MVD} generalize the concept of FDs, since every FD is an \gls{MVD}, but not every \gls{MVD} is an FD. 
% In simplest form, if we consider only two attributes, an \gls{MVD} allows multiple values of one attribute to be linked to the one value of another attribute. 
% When this association is limited to exactly one value, the constraint reduces to an FD. 

% Therefore, FDs can be viewed as a special case of \glspl{MVD}.
}
% ----- Definition ------------------------------------------------------------
{
\label{def:fd-violation}
Let $A_{lhs} \subseteq A$ be the non-empty sets of attributes in a relation $r$ and $a \notin A_{lhs}$ be another attribute in $r$, where an FD is defined as $A_{lhs} \to a$. 
A functional dependency violation occurs if there exist two distinct tuples $t_1, t_2$ in $r$, such that $t_1.A_{lhs}.v = t_2.A_{lhs}.v \;\land\; t_1.a.v \neq t_2.a.v$. 
}
% ----- Example ---------------------------------------------------------------
{
An FD is defined on the \schema{Employee} relation as \schema{EID} $\to$ \schema{Email}, implying that if two tuples have the same employee ID, then they should have the same Email. So, the tuples e3 and e8 in \schema{Employee} relation break the FD.
}

% ----- CONDITIONAL FUNCTIONAL DEPENDENCY VIOLATION -----
\error[err:cfd-violation]
{Conditional functional dependency (CFD) violation}
% ----- Description ------------------------------------------------------------
{
CFDs are a generalization of traditional FDs. 
While FDs were developed primarily for schema design and to enforce normalization, CFDs broaden their role to data quality by enforcing consistency across semantically related values~\cite{Bohannon2007, Golab2008, Fan2008, Fan2012}. A simple case of CFD can be illustrated as $a_1 \to a_2$ in a relation, with an additional condition on another attribute as $a_3 = c$. 
This implies that whenever two tuples agree on the value of attribute $a_1$ and satisfy the condition $a_3 = c$, then they must also agree on the value of attribute $a_2$.
Unlike standard FDs, the CFDs hold only under certain conditions, and when data does not adhere to these conditions, a CFD violation occurs~\cite{Josko2016}.
}
% ----- Definition ------------------------------------------------------------
{
\label{def:cfd-violation}
Let $A_{lhs} \subset A$ and $a, a_{cond} \in A$, where $A_{lhs}$ is a non-empty set of attributes in a relation $r$ and $a, a_{cond} \notin A_{lhs}$.
A CFD is defined as $A_{lhs} \to a$, if $a_{cond}$ matches a predefined condition $c$. 
A CFD violation occurs in the relation $r$ if there exist two tuples $t_1$ and $t_2$ such that $(t_1.a_{cond} = t_2.a_{cond} = c) \;\land\; (t_1.A_{lhs} = t_2.A_{lhs}) \;\land\; (t_1.a \neq t_2.a)$.

% 2. Let $a_1$, $a_2$ and $a_3$ be three attributes in relation $r$, where a conditional functional dependency is defined as $a_1 \to a_2$, given $a_3 = c$. A conditional functional dependency (CFD) violation occurs if there exist two tuples $t_1, t_2 \in r$ such that $t_1.a_3 = t_2.a_3 = c \;\land\; t_1.a_1 = t_2.a_1 \;\land\; t_1.a_2 \neq t_2.a_2$.
}
% ----- Example ---------------------------------------------------------------
{
We can define a CFD as \schema{EID} $\to$ \schema{FullName}, with the additional condition that the \schema{DID} corresponds to the \data{IT} department in \schema{Employee} relation. 
So, the tuples e1 and e9 violate the defined CFD due to the inconsistency in the full name, even though both tuples have the same employee ID and belong to the IT department.
}

% ----- CYCLIC DEPENDENCY VIOLATION -----
\error[err:cyclic-dep-violation]
{Cyclic dependency violation}
% ----- Description ------------------------------------------------------------
{
A cyclic dependency violation occurs when two or more tuples in a self-referencing relation form a loop. Oliveira et al.~\cite{Oliveira2005} refer to such violations as circularity among tuples in a self-relationship, where starting from a tuple and following its foreign key references, the cycle leads back to the same tuple. The result of such circular dependency among tuples may violate the semantics of a data model that is expected to be acyclic.
}
% ----- Definition ------------------------------------------------------------
{
% \label{def:cyclic-dep-violation} Let $a_1$ and $a_2$ be two attributes in a relation $r$, such that $a_1$ determines $a_2$ and $a_2$ determines $a_1$. These constraints imply a one-to-one correspondence between attributes, $a_1$ and $a_2$. A cyclic dependency violation occurs if there exist two tuples $t_1, t_2 \in r$ such that $(t_1.a_1.v = t_2.a_1.v \; \land \; t_1.a_2.v \neq t_2.a_2.v) \;\lor\; (t_1.a_2.v = t_2.a_2.v \; \land \; t_1.a_1.v \neq t_2.a_1.v)$.

\label{def:cyclic-dep-violation}
Let $r$ be a self-referencing relation with attributes $a_{PK}, a_{FK} \in A$, where $a_{PK}$ is the primary key and $a_{FK}$ is a foreign key referencing $a_{PK}$. A cyclic dependency violation occurs if there exist tuples $t_1, t_2, \dots, t_n \in r$, with $n \geq 1$, such that $(t_i.a_{FK}.v = t_{i+1}.a_{PK}.v$ for $1 \leq i < n)$ and $(t_n.a_{FK}.v = t_1.a_{PK}.v)$.
}
% ----- Example ---------------------------------------------------------------
{
A cyclic dependency can be observed in the \schema{Employee} relation as tuple e1 with \schema{EID} = 101 has \schema{MID} = $102$ and tuple e2 with \schema{EID} = 102 has \schema{MID} = $101$. This is a cyclic dependency violation: the \schema{MID} references lead from $e_1$ to $e_2$ and back to $e_1$, forming a cycle.
}

% ----- BUSINESS RULE VIOLATION -----
\error[err:rule-violations]
{Rule violations}
% ----- Description ------------------------------------------------------------
{
Rule violations occur when the data fails to satisfy predefined rules. 
Any violations of business regulations, legal authority rules, or database administrator rules are considered rule violations.
These rules are not limited to predefined constraints; they can also include expected properties of the data derived from domain knowledge or metadata. 
Rule violations can occur at all granularity levels, affect both syntactic and semantic contexts of data, and can be caused by specific actions or inactions on the data.

\textit{Business rule violation}: A business rule is violated when the data fails to comply with the policies and regulations defined in a business context~\cite{Oliveira2005}. Business rules are typically defined internally at the enterprise level.
The Semantics of Business Vocabulary and Business Rules (SBVR)~\cite{OMGSBVR2019}, introduced by the Object Management Group (OMG), provides a guideline for the standardized definition of such rules across and within organizations.
Instead of providing a fixed set of business rules, SBVR suggests how organizations can define their own rules (expressed using business vocabulary~\cite{Ashish2011}) in a standard way, which can aid in reducing ambiguity and misinterpretation across departments in large organizations.
One specific case of a business rule violation occurs when an organization's human resources system requires every employee record to include a department identifier. In contrast, another internal analytics system does not enforce this requirement. When someone prepares a dashboard and imports employee data from that system, the resulting records lack departmental information, violating a business rule~\cite {Minock2015}.

\textit{Database administrator (DBA) rule violation}: A DBA~\cite{Tamer2018} is responsible for defining rules and guidelines related to data management, storage, and access of one or multiple information systems. A DBA rule violation occurs when data fails to comply with the constraints set by the database administrator, which often leads to errors or inconsistencies~\cite{Ge2007ARO}. Ge and Helfert~\cite{Ge2007ARO} refer to such violations by the term \textit{Violation of constraints provided by the database administrator}, but in this catalog, we refer to them by the term DBA rule violations.

\textit{Legal rule violation}: A legal rule violation, also referred to as \textit{Violation of Company and Government Regulations}~\cite{Ge2007ARO}, occurs when data does not comply with the regulations, standards, and constraints set by legal entities or governing bodies. An example of widely recognized government regulation is the Sarbanes-Oxley Act~\cite{law2002}, a U.S.-based regulation that gives guidelines to improve transparency in the financial reporting of public companies and accounting firms. 
Beyond financial compliance, legal regulations are increasingly used to protect personal and sensitive information in the massive volumes of data generated every day.
The General Data Protection Regulation (GDPR) is one such legal regulation introduced by the European Union (EU) to protect the personal data of its citizens and ensure that organizations handling the data of individuals in the EU comply with its provisions, regardless of their location~\cite {Cejas2023}. Such regulations also apply to the data handled by many widely used applications, such as Alexa and Siri, to ensure the responsible use and secure storage of user information.
}
% ----- Definition ------------------------------------------------------------
{
\label{def:rule-violations} Let $e_{rules}$ be a set of all the database elements (values, tuples, attributes, and relations) in an RDB to which one or more business, DBA, or legal rules apply. 
Let $CheckRules(e)$ be a boolean function that returns $\true$ if all rules defined for a database element $e \in e_{rules}$ are satisfied. A rule violation occurs when $CheckRules(e) = \false$, indicating that the corresponding element does not satisfy the defined rules. 
}
% ----- Example ---------------------------------------------------------------
{
The following are examples of various rule violations:

\begin{itemize}
  \item \textbf{Business rule violation:} A business rule violation occurs when an organization enforces a rule that every employee is assigned to precisely one department, reflecting the real-world structure of roles and responsibilities within the company. If an employee is mistakenly linked to multiple departments, the record violates this business rule. 
  \item \textbf{DBA rule violation:} A DBA rule requires a standard naming structure that requires the \schema{FullName} attribute to contain at least two words with the first name listed first (mandatory), followed by an optional middle name, and then the last name without any comma separator. 
  A DBA rule violation can be observed in tuple e3 as the full name contains commas, which breaks the enforced formatting convention. While this may appear to be a formatting error, it is classified here as a business rule violation because the requirement is defined by an organizational rule.
  \item \textbf{Legal rule violation} A legal rule violation can occur when employees' data that should be collected or processed only with explicit user consent is stored or used outside the allowed usage. For instance, under the GDPR, personal data collected from employees for emergency contact cannot be repurposed for targeted advertising without additional consent. If the system stores its data solely for emergency contact purposes and that same data later appears in a marketing dataset without evidence of consent, the company violates the legal requirements for lawful storage and processing of the data.
\end{itemize} 

}

\error[err:outdated-data]
{Outdated data}
% ----- Description ------------------------------------------------------------
{
Outdated data refers to any tuple that once reflected correct information but no longer represents the current state of its real-world origins~\cite{Ge2007ARO}. It can result from a change in the properties of a real-world element or a relationship represented in an RDB. For example, the address currently stored for an employee becomes outdated if not updated after they relocate. 
According to Lee et al.~\cite{Lee2006}, even when the underlying schema remains the same, data that once met user needs may become outdated if those needs change. For instance, when new regulations introduce updated certification standards, valid certification records of multiple employees may become outdated.
Outdated data is a form of temporal obsolescence that can occur even in the absence of explicit temporal attributes, such as timestamps or validity intervals. 
Readers seeking a focused discussion of errors that depend directly on temporal attributes can refer to the taxonomy of temporal data defects by Josko et al.~\cite{Josko2018}.
}
% ----- Definition ------------------------------------------------------------
{
\label{def:outdated-data} Let $M^{\tau}(t)$ be the mapping that returns the real-world state represented by tuple $t$ at the current time $\tau > \tau'$, then $t$ will be considered outdated if $t.v = M^{\tau'}(t) \;\land\; t.v \neq M^{\tau}(t)$.

% Let a tuple $t^{\tau}$ be valid with respect to the current schema of the relation $r^{\tau}$ at a given point of time $\tau$. With $\tau'$ > $\tau$, the schema has changed to $r^{\tau'}$ and the tuple $t^{\tau}$ is considered outdated if $t^{\tau} \not\models r^{\tau'}$.
}
% ----- Example ---------------------------------------------------------------
{
The salary of the employee in tuple e1 was increased by ten percent, but the corresponding tuple in the \schema{Employee} relation still stores the old salary. As the database was not updated to reflect the change in the real-world value, the tuple now contains outdated data.
}

\section{Redundant Data}
\label{sec:redundantdata}
\glsreset{RDB}
\glsreset{FD}
\glsreset{DQ}

Data in an RDB are considered \emph{redundant} when information is either duplicated or stored in a form that is unnecessary, as it can be derived from other attributes, tuples, or relations in the database. 
New Oxford American Dictionary~\cite{NOAD} defines the term \emph{redundant} as ``not or no longer needed or useful'', highlighting that redundancy includes more than duplication.
In the context of data management, redundant data refers to any additional information that adds no value or whose removal would not affect the intended task, resulting in anomalies in update, insertion, and deletion tasks, as well as redundant storage~\cite{Ramakrishnan2003}.
Understanding the different forms of redundancy is important because redundant data can create the illusion of having more information than is actually present~\cite{Talburt2013} or can cause us to overlook missing information that the system truly needs. 
For example, if management wants to select a cohort of 200 employees from the city of Berlin, redundant records create the illusion of complete data, but in fact, some records are missing.
Errors resulting from redundant data can be broadly categorized into two categories based on their causes. The first category is intra-source duplicates, which are unintentional repeated entries of the same record in a database. 
The second category is inter-source duplicates, which get introduced in a database during the data integration process from multiple sources~\cite{Lee2006, Naumann2010}. 
Efforts to identify and resolve data redundancy span several research communities under different names. 
In the statistics and data science fields, this work is often referred to as ``record linkage'' and ``record matching''. At the same time, the database community commonly uses terms such as ``database hardening'', ``name matching'', ``merge-purge'', and ``de-duplication''~\cite{Talburt2013, Fan2012}.

\error[err:duplicate-tuples]
{Duplicate tuple}
% ----- Description ------------------------------------------------------------
{
A tuple in a relation is considered a duplicate if two or more tuples refer to the same real-world entity, regardless of whether their attribute values are identical, partly similar, or even substantially different due to alternate formats, partial data, or missing information~\cite{Elmagarmid2007, Naumann2010, Fan2012, Fan2015, Josko2016}. For example, the (fake) phone numbers %\footnote{\url{https://fakenumber.org/}} 
\data{+1-561-555-7689} and \data{561-555-7689} represent the same phone number, with the first including the country code. Such heterogeneous representation of the same value is referred to as lexical heterogeneity~\cite{Elmagarmid2007} or fuzzy duplicates~\cite{Naumann2010}. 
Duplicate tuples are also studied in the literature as part of a broader problem known as ``entity identity integrity''. According to Talburt and Zhou~\cite{Talburt2013}, entity identity integrity requires that each real-world entity be represented by exactly one tuple and that no two distinct entities share the same representation.
}
% ----- Definition ------------------------------------------------------------
{
\label{def:duplicate-tuple} All the tuples $t_i \in T_{dup} \subseteq T$ in a relation $r$ are considered duplicates if $ \; \exists \allowbreak \text{ a real world entity} \; m \;\text{such that } \forall \; r.t_i \in T_{dup},\;M(r.t_i) = m$.
}
% ----- Example ---------------------------------------------------------------
{
Tuples c2 and c3 in the \schema{Certificate} relation describe the same certificate, which makes them duplicate tuples.
}

% Irrelevant row
\error[err:irrelevant-row]
{Irrelevant data}
% ----- Description ------------------------------------------------------------
{
Irrelevant data occurs when a database tuple does not represent any information about a real-world entity or its relationships that the database is supposed to contain. 
Such data can be valid and correctly formatted, but still cause problems for downstream tasks because the data in these tuples is not the one that the relation is supposed to represent~\cite{Josko2016}. 
}
% ----- Definition ------------------------------------------------------------
{
\label{def:irrelevant-data}A tuple $t$ in a relation $r$ is considered irrelevant if $r.t \notin M(r)$.
}
% ----- Example ---------------------------------------------------------------
{
The tuple c4 in the \schema{Certificate} relation should not be included, as the company records only SAP certifications, and the Azure certification does not belong in this relation.
}

\section{Beyond Data-Level Error Types}
% Optional: Additional Data Quality Factors
\label{sec:beyond-error-types}
\glsreset{RDB}
\glsreset{FD}
\glsreset{DQ}

While related surveys on data errors (cf. \autoref{sec:related-work}) often blur the distinction between actual data errors, data characteristics that may lead to errors, and metadata errors, this paper maintains a clear focus on data-level errors. This section explicitly and briefly addresses metadata errors (\autoref{sec:metadata}) and related data characteristics (\autoref{sec:related-data-characteristics}) to clarify these boundaries.

\subsection{Metadata Errors}
\label{sec:metadata}
Metadata is commonly described as ``data that defines or describes other data''~\cite{ISO8000}, and it provides information about the nature, structure, storage, and management of data in databases. 
In the context of this survey, metadata refers to all schema elements of an RDB, including relation and attribute labels, data types, constraints, and database statistics.
In essence, the various data errors discussed in this survey can likewise appear in metadata. Examples include entire duplicate relations, duplicate or missing attributes, missing constraints, unnecessary constraints, or misspelled attribute labels. 
These errors often have their roots in data integration scenarios (cf.~\cite{Batista_2007, DuchateauBellahense2010}) and can directly affect the quality of data in a database~\cite{Ehrlinger2018}. 
For example, erroneously enforcing a \schema{NOT NULL} constraint on a \schema{MiddleName} attribute may lead to the insertion of disguised missing values (DMVs) (cf.~\autoref{err:disguised-missing-val}). 
Vice versa, the absence of constraints on mandatory attributes, such as \schema{ID} or \schema{DateOfBirth}, can result in records with missing identifiers and incomplete information. 
Such missing IDs can further result in duplicate tuples and cause record linkage issues.

High-quality metadata is therefore a prerequisite for high-quality data~\cite{Quarati2023, Cima2025}. 
Although metadata errors are an important topic, a comprehensive investigation is beyond the scope of this survey.
We encourage readers to consult the metadata literature alongside this survey, e.g.,~\cite{Batista_2007, DuchateauBellahense2010, Ehrlinger2018b} to better understand metadata quality, which can help prevent many data-level errors.

\subsection{Related Data Characteristics}
\label{sec:related-data-characteristics} 

We have collected related data characteristics from the literature, which do not signify data errors; however, they can cause various data errors or prevent data from being used for its intended purpose.
These related data characteristics either require domain knowledge or even require knowledge beyond the domain for their detection.
For instance, the issue of \et{information not being based on facts}~\cite{Ge2007ARO} is domain-dependent, as it requires understanding the domain to detect it.
On the other hand, the issue of \et{information that is challenging to aggregate}~\cite{Ge2007ARO} is domain-independent and arises when an expected data aggregation (e.g., for salaries or sales counts) across tuples or attributes fails to yield meaningful results. 

These issues become more challenging when one issue leads to another. 
For instance, the case when \et{the information is inaccessible}~\cite{Ge2007ARO} encompasses both the availability and retrievability of data.
Information is considered inaccessible when both the data and its associated metadata are irretrievable, thereby limiting their availability for use~\cite{Wang1996, Hassenstein2022}. 
Therefore, irretrievability can be one cause of inaccessible information, but inaccessibility can also result from many other reasons, such as access restrictions or policy constraints. 
Thus, such issues, which can result from multiple factors, make maintaining overall data quality particularly challenging.
The following are some of the data-related characteristics and issues reported in the literature:

In particular, \citet{Ge2007ARO} list the following characteristics about information that we do not consider to be data errors: the information is difficult to aggregate, inaccessible, hardly retrievable, insecure, not based on facts, of doubtful credibility, an impartial view, is irrelevant to the work,  is compactly represented, hard to manipulate, hard to understand, or consists of inconsistent meanings.

% \begin{itemize}
%     \item User-/information-level quality issues from~\citet{Ge2007ARO}:
%     \begin{itemize}
%         \item The information is difficult to aggregate.
%         \item The information is inaccessible.
%         \item The information is hardly retrievable.
%         \item The information is insecure.
%         \item The information is not based on facts.
%          \item The information is of doubtful credibility.
%         \item The information presents an impartial view.
%         \item The information is irrelevant to the work.
%         \item The information consists of inconsistent meanings.
%         \item The information is compactly represented.
%         \item The information is hard to manipulate.
%         \item The information is hard to understand~\cite{Ge2007ARO} or an embedded value~\cite{RahmDo2000}
%     \end{itemize}

%     \item Database integrity guaranteed through transaction management from Kim et al.~\cite{Kim2003}:
%         \begin{itemize}
%             \item Lost update (due to lack of concurrency control)
%             \item Dirty read (due to lack of concurrency control)
%             \item Unrepeatable read (due to lack of concurrency control)
%             \item Lost transaction (due to lack of proper crash recovery)
%         \end{itemize}
% \end{itemize}

These issues can affect how data is collected, stored, accessed, and processed. 
% These data-related characteristics are very clear from their names themselves. 
Therefore, it is important to consider them when working with data. They should not be confused with the data error types discussed in previous sections. 
They highlight how data-related characteristics can introduce limitations to data usability, even in the absence of data errors, and can act as blockers or amplifiers of data errors rather than errors themselves. 
% \input{sections/7_metadata}
%\section{Data Error Management}
\section{Detecting and Correcting Data Errors} 
\label{sec:errormanagment}
\glsreset{DQ}

%Even after high-quality metadata preparation, the data may contain errors. 
%%LISA: I commented out all those parts because they center very much around data management and might not be focused enough for this section
%To understand, identify, and correct data errors, a set of practices has been proposed in the data cleaning literature~\cite{Ilyas_2019,Ganti_2013}
% These activities include elicitation, planning, measurement, diagnosis, treatment, and maintenance. 
% Moreover, \gls{DQ} criteria are often ill-defined, have missing metadata, and may require expensive quality assessment, a challenge that has also been reported in applied domains such as clinical research by Van den Broeck et al.~\cite{VanDenBroeck2005}.
%Improving \gls{DQ} for various downstream tasks, such as those involving ML models, still remains challenging~\cite{Yuhan2024}.

Data error handling relies on two closely related activities: \emph{error detection} and \emph{error correction}. 
These activities are essential parts of data preparation pipelines, and are often discussed under the broader umbrella of \emph{data cleaning}~\cite{Ilyas_2019, Ilyas_2015, Zhu2025, Ganti_2013}.
These two activities have inspired the development of various tools to detect and correct (repair) data errors~\cite{Ehrlinger_2022, Yuhan2024, Papastergios_2025}. 
This section briefly discusses methods and tools for data error detection (\autoref{sec:error-detect}) and correction (\autoref{sec:error-correct}) that practitioners can use for data cleaning. 
In addition, \emph{error prevention} strategies (such as enforcing integrity constraints and formatting requirements) are essential to prevent future errors from entering the system.
However, real-world data systems can continue to accumulate errors even after error prevention strategies due to evolving requirements, heterogeneous data sources, and human or software issues~\cite{Schelter_2018b}. 
For a comprehensive discussion on data cleaning, we refer to \cite{Ganti_2013} and \cite{Ilyas_2019}. 

\subsection{Detecting Data Errors}
\label{sec:error-detect}

Error detection identifies violations of expected data properties. 
Methods for error detection include both qualitative techniques, which are based on rules, constraints, and patterns, as well as quantitative techniques based on statistical deviations~\cite{Chu2016DC, Chu2016QDC, Abedjan_2016, Ding2022}.
Along with these, error detection can also be viewed from a DQ dimensions perspective, as different types of data errors affect different DQ dimensions and their metrics.
For example, \et{missing values} primarily affect completeness, whereas violations of \et{integrity constraints} indicate a lack of consistency. 
However, it remains an open challenge to establish a clear and comprehensive mapping between data errors, DQ dimensions, and their corresponding metrics.

In qualitative error detection~\cite{RahmDo2000, Chu2016DC, Fan2008DC}, the ability to detect errors depends not only on the data itself, but also on external contextual information, which is often incomplete, missing, outdated, or inaccessible (see \autoref{sec:metadata} for metadata errors). 
Modern error detection approaches address these challenges by combining signals from both metadata and data and typically identify potential errors rather than their root causes. 
For example, a \et{spelling mistake} could result from a typographical error or a semantic mismatch.
Error source identification becomes more challenging when multiple mechanisms can introduce the same error into the data. 
For instance, even after detecting data was missing during data entry, an error detection method still needs to determine whether the data is missing completely at random (MCAR), missing at random (MAR), or missing not at random (MNAR)~\cite{Little_1983}. 
Understanding these underlying mechanisms is crucial for identifying root causes and implementing future error prevention strategies.

Almost all error detection methods are computationally complex and require some level of human expertise for guidance. 
Example tools for error detection are Holodetect~\cite{Heidari_2019}, which uses neural networks for weakly supervised error detection, DQ-MeeRKat~\cite{Ehrlinger_2021b}, which employs explainable statistics to learn dataset-specific data profiles, Deequ~\cite{Schelter_2018b}, which enables user-defined quality rules and ML-enhanced constraint suggestions, and Raha~\cite{mahdavi2019raha}, currently the most effective general-purpose error detection tool across benchmarks~\cite{Ni_2024}.
Large language models (LLMs) are an emerging direction, offering a significant advancement in error detection due to their extensive contextual understanding.
They allow the detection of overlooked errors as well as the suggestion of cleaning strategies as further discussed in the next section~\cite{Zhang_2025}.

\subsection{Correcting Data Errors}
\label{sec:error-correct}
Error correction, also referred to as data repair, addresses how detected errors are corrected to restore or improve DQ~\cite{Ding2022}. 
Typical data correction activities include imputing missing values, modifying data values (e.g., to adhere to a given format), or merging duplicate tuples. 
These activities can be performed with a wide range of methods, ranging from  traditional statistical and constraint-based methods to ML and DL techniques~\cite{Zhu2025}. 
The choice of the respective error correction method depends on assumptions about the origin of an error.
% on uimputing missing values, modifying data values (e.g., to adhere to a given format), or merging ta itself, the constraints on the data, or both are erroneous~\cite{Chu2016DC}. 
In many cases, detected errors can also have multiple plausible repairs. Thus, data correction procedures often require additional context to decide on the optimal repair action or enumerate all violations in the data to derive a minimal repair~\cite{Rekatsinas_2017}.
However, data correction requires careful analysis beforehand, as it can introduce bias or distort data distributions if not done thoroughly~\cite{Khayyat2015}.

\section{Related Work}
\label{sec:related-work}
%\glsreset{DQ}

Our survey formalizes a superset of data errors and error indicators that have already been discussed in the literature. 
While prior work subsumes both categories under the general term ``data error types'', we highlight the difference as useful information for handling data errors.

Prior attempts to categorize error types acknowledge the need for such a catalog, but differ significantly in scope. 
These works cover only parts of the data errors and error indicators discussed in our catalog.
Moreover, most of them do not provide formal definitions (except for \cite{Oliveira2005, Josko2016}), and the classification criteria are not unified. 
This fragmentation of error types motivates our effort to consolidate the current landscape. 
Hence, we reconciled all existing data error categorizations and expanded the scope of our survey to include previously excluded errors. 
In this section, we provide a brief review of related work on error types classification and definition, and distinguish it from our survey, in increasing order of publication year.

The influential work by Rahm and Do~\cite{RahmDo2000} classifies 14 single-source data error types (10 value-level errors and four constraint violations), but does not provide formal definitions or clear distinctions. 
% that affect individual values or violate schema-level constraints within a relation. The single-source problems include 
Examples include missing or illegal values, duplicate tuples, or attribute dependency violations.
For multi-source problems, Rahm and Do~\cite{RahmDo2000} do not introduce new error types but highlight that single-source errors can be aggravated when data from multiple sources are integrated. 
Even with this limited number of error types, Rahm and Do~\cite{RahmDo2000} have had a significant impact as one of the earliest works in data error classification, clearly demonstrating the need to consolidate the scattered landscape of data errors.  

Müller and Freytag~\cite{Mueller2003} discuss nine error types (referred to as anomalies) in the context of data cleansing, categorized as \textit{syntactic}, \textit{semantic}, and \textit{coverage} anomalies.
%and provide a high-level conceptual view of common \gls{DQ} problems. 
However, their categorization remains high-level, with no detailed descriptions of the error types, leaving readers unclear how they manifest in actual data. 
For example, one of the nine error types in~\cite{Mueller2003} is \et{lexical errors}, which can subsume multiple error types such as misspellings, heterogeneous formatting issues, or syntax violations. 
Similarly, error type \textit{irregularities} can also subsume multiple error types, such as outliers or incorrect units, without explaining their differences or potential overlaps. 
In contrast, our catalog formalizes these specific error types in detail.  

Kim et al.~\cite{Kim2003} classify 33 error types into the three categories \textit{missing}, \textit{not missing but wrong}, and \textit{unusable}, based on error manifestation and using a hierarchical refinement approach.
However, this classification primarily provides error names without sufficiently describing their nature, definitions, potential overlaps, or root causes. 
In addition, several error types (e.g., \et{lost update}, \et{dirty read}, \et{lost transaction}) represent system-level issues that can lead to errors but do not describe specific error types themselves. 
We consider these issues to be a better fit as related data characteristics and therefore mentioned them in \autoref{sec:related-data-characteristics}.
While~\cite{Kim2003} is helpful as a list of error type names, it does not constitute a well-founded taxonomy of data error types.

Oliveira et al.~\cite{Oliveira2005, Oliveira2005b} categorize data error types according to the database granularity level at which they are detected -- 35 error types in \cite{Oliveira2005b} and 30 in \cite{Oliveira2005}.
While both works provide textual descriptions and illustrative examples, only \cite{Oliveira2005} includes formal definitions, making it the closest prior work to ours. 
%% From here! 
%% Differentiate from our works
% The error taxonomy in \cite{Oliveira2005b} is comprehensive and even without definition very useful; however, some of the error types are symptom-oriented. 
Despite this comprehensive approach, both taxonomies have specific issues that limit their practical use.
In both taxonomies, the classification by granularity and data source type (single-source vs. multi-source) creates redundancy. 
For instance, in \cite{Oliveira2005b}, error type \et{domain constraint violations} appear 6 times, indicating that the 35 listed errors do not represent 35 unique error types. 
Similarly, error type \et{violation of business domain constraint} in \cite{Oliveira2005} appears 4 times under different granularity levels.
In addition, some categories are symptom-oriented; for instance, the category \et{meaningless value} in \cite{Oliveira2005b} describes how errors appear in the data rather than their specific root cause.
A meaningless value can result from different underlying problems, including disguised missing values, encoding errors, or data corruption, which are not distinguished in their work. 
Therefore, it is not clear into which error category (missing, incorrect, or redundant) the meaningless value error type should be placed.

Ge et al.~\cite{Ge2007ARO} classify 29 errors according to (i) data versus user perspective, and (ii)  context-dependent versus context-independent perspective. 
Of these 29 errors, 17 are data error types (included in our categorization), while the remaining 12 belong to related data error characteristics discussed in  \autoref{sec:related-data-characteristics}. 

Josko et al.~\cite{Josko2016} classify 26 data error types based on the \gls{DQ} dimensions \emph{accuracy}, \emph{completeness}, and \emph{consistency}. 
The three-step methodology they introduced distinguishes between constraint violations and fact representation deviations and categorizes errors as dynamic versus static and explicit versus implicit. 
While \cite{Josko2016} captures important characteristics of data error types and their occurrence levels, it does not cover all error types covered in previous surveys or across all data granularity levels, such as semi-empty rows or tuples. 
Note that Josko also provides a formal definition of temporal data errors in \cite{Josko2018}, which constitutes a complementary extension to our work and discusses the concept of outdated data (\autoref{def:outdated-data}) in detail. 

Ilyas and Naumann~\cite{Ilyas2022} do not propose a fine-grained taxonomy of concrete data error types, but instead discuss a small number of high-level categories based on error symptoms. These error categories are outliers, missing values, constraint violations, and duplicates.
Their work emphasizes the need for research into the root causes and provenance of error types in data pipelines rather than on cataloging specific error types.

Our survey is a unified reference for \gls{DQ} research from a data-error perspective.
We include many error types directly from related work and generalize others, describing their subtypes in the corresponding error descriptions.
For instance, under constraint violation (see \autoref{err:const-violation}), we list multiple subtypes, including uniqueness, referential integrity, recursive, transition, and temporal constraint violations. 
In \autoref{tab:error_org}, we provide an overview of error types that are discussed for the first time in a taxonomy by us, those that are included both in our work and in prior surveys, and those that we do not directly discuss, together with the reasons listed in the table.
Consequently, we created a joint list from all these surveys. 
We extended it in two directions: (i) we identified subtypes and variants of the data error types discussed in these works, and (ii) we added newly emerged errors. The fact that a few error types were not included or were only partially discussed in the older surveys (such as disguised missing values, invalid values, and out-of-vocabulary values) further underscores the need for a refreshed, comprehensive survey. 
\autoref{tab:error_org} provides an overview of all error types collected from related work, a justification for error types that we did not include, as well as a list of all error types we newly added that were not covered by previous classifications. 

\begin{table}
    \centering
    \begin{adjustbox}{width=1\textwidth}
    \scriptsize
    \begin{tabular}{p{0.23\textwidth}|p{0.33\textwidth}|p{0.45\textwidth}} \toprule
    Our Work Only & Our Work and Related Error Classifications& Discussed in Related Error Classifications\\ \midrule
    Disguised Missing Value & Missing Values \cite{RahmDo2000, Mueller2003, Kim2003, Oliveira2005, Josko2016, Ilyas2022} &  Referential Integrity Violation~\cite{RahmDo2000, Oliveira2005}, which is mentioned as subtype of constraint violation (\autoref{def:const-violation})
    \\ \vspace{0.2cm}
    Empty Attribute & Partial-Empty Tuple/Attribute \cite{Oliveira2005} & Embedded Value~\cite{RahmDo2000}, which is mentioned in ``related data characteristics'' 
    (\autoref{sec:related-data-characteristics}) since it must not necessarily be an error
    \\ \vspace{0.2cm}
    Biased Data & Missing Tuple \cite{Mueller2003, Josko2016} & Lexical errors \cite{Mueller2003}, which can belong to multiple data error categories (structural / format-level)
    \\
    Out-of-Vocabulary (OOV) Word & Invalid Value/Tuple \cite{Oliveira2005, RahmDo2000, Mueller2003} & Irregularities~\cite{Mueller2003}, which fits better with metadata/representation-level inconsistency
    \\
    Typo & Misspelling \cite{RahmDo2000, Oliveira2005} & Existence of homonyms/homonymous values ~\cite{Oliveira2005, RahmDo2000, Josko2016}, are mentioned under semantically ambiguous data data error category (\autoref{def:semantic-amb})
    \\
    Misscan & Synonym \cite{Oliveira2005, Josko2016, RahmDo2000} & Heterogeneity of measure units \cite{Oliveira2005}, fits better under metadata/ representation-level inconsistency 
    \\
    Incorrect Encoding & Word Transposition \cite{RahmDo2000} & Heterogeneity of representation \cite{Oliveira2005}, fits under metadata/representation-level heterogeneity
    \\
    Noise & Misfielded Value \cite{RahmDo2000} 
    & Transition Constraint Violation~\cite{Josko2016}, mentioned under constraint violations (\autoref{def:const-violation})
    \\
    Incorrect Unit & Semantically Ambiguous Data \cite{Oliveira2005} & Atypical Tuple~\cite{Josko2016} is a symptom-based anomaly (error indicator, not a single error type)
    \\
    Partially empty attribute & Syntax Violation \cite{Oliveira2005} &  Cardinality Ratio Violation~\cite{Josko2016}, included under constraint violations (\autoref{def:const-violation})
     \\
    & Heterogeneous Formatting \cite{Oliveira2005} & Disjoint Subdomains~\cite{Josko2016}, mentioned under functional dependency (FD) violations (\autoref{def:fd-violation})
    \\
    & Incorrect Reference \cite{RahmDo2000, Oliveira2005, Josko2016} & Invalid Tuple~\cite{Josko2016} is included under invalid tuples (\autoref{def:invalid})
    \\
    & Constraint Violation \cite{Ge2007ARO, Mueller2003, Kim2003, Ilyas2022} & Heterogeneous granularity~\cite{Josko2016} is included under incorrect unit (\autoref{def:incorrect-unit})
    \\
    & Domain Constraint Violation \cite{Josko2016, Ge2007ARO, Oliveira2005} & Incompatible Replication \cite{Josko2016} is a FD-like consistency rule due to heterogeneous representation of redundant copies (not a single error type)
    \\
    & Uniqueness Violation \cite{Ge2007ARO, Oliveira2005, RahmDo2000, Kim2003} & 
    \\
    & Attribute Dependency Violation \cite{RahmDo2000} & \\
    & Functional Dependency (FD) Violation \cite{Oliveira2005, Ge2007ARO, Josko2016} & 
    \\
    & Conditional FD Violation \cite{Josko2016} & 
    \\
    & Cyclic Dependency Violation \cite{Oliveira2005} & 
    \\
    & Business Rules Violation \cite{Oliveira2005, Ge2007ARO} & \\
    & Legal Rules Violation \cite{Ge2007ARO} & 
    \\
    & DBA Constraint Violation \cite{Ge2007ARO} & 
    \\
    & Outdated Data \cite{Ge2007ARO, Kim2003} &  
    \\
    & Duplicate Tuples \cite{RahmDo2000, Mueller2003, Oliveira2005, Ge2007ARO, Josko2016, Kim2003, Ilyas2022} & 
    \\
    & Irrelevant Data \cite{Ge2007ARO} & 
    \\
    & Outlier \cite{Ilyas2022} & \\
    \bottomrule
    \end{tabular}
    \end{adjustbox}
    \caption{Error types discussed (1)~in our work, (2)~both in our work and in related data error classifications, as well as (3)~only in related classifications explicitly including a explanation how these error types are implicitly covered in our work \cite{RahmDo2000, Mueller2003, Oliveira2005, Ge2007ARO, Josko2016, Kim2003, Ilyas2022}.}
    \label{tab:error_org}
\end{table}
\section{Conclusion}
\label{sec:conclusion}
\glsreset{DQ}

%% What have we done
This survey provides a comprehensive and formal catalog of \toterrornum data error types and error indicators, which are classified into three non-overlapping categories: \emph{missing}, \emph{incorrect}, and \emph{redundant} data, based on their manifestations in tabular datasets. 
For each error type, we provide a formal definition, practical examples, and references to relevant literature to clarify and align terminology.
The selection of data errors is based on (1)~a consolidation of existing data error taxonomies \cite{RahmDo2000, Mueller2003, Oliveira2005, Ge2007ARO, Josko2016, Kim2003, Ilyas2022}, for which we highlighted and resolved inconsistencies between them, as well as (2)~an extension of these works with underexplored error types that are specifically relevant for artificial intelligence use-cases and were not covered by existing data error taxonomies. 
%The current data quality literature often exhibits terminological ambiguity, with similar error types addressed using different names across taxonomies. 

%% Why is our useful? highlight benefits again! 
The catalog of data errors serves both practical and research purposes. It enables practitioners to implement error-specific detection and cleaning strategies and supports researchers in investigating underexplored error types that lack adequate detection methods and tools. 
For future research, we envision several important directions: 

\textbf{Automated data error classification.} Several studies show that knowledge of the specific error type can inform and improve existing error detection and cleaning approaches~\cite{Schelter2021, mohammed2025step}. Yet, no current error detection tool returns \emph{the type of error} it detected.
Thus, as a direct follow-up to this work, we plan to develop new approaches to classify and thereby differentiate specific data error types. 
% We deem knowledge about the specific error type as extremely important to inform and improve current error detection and cleaning strategies. 

\textbf{Formalization of metadata errors.} While errors in the metadata can directly cause data-level errors, there is no comprehensive classification or formalization for these metadata errors. 
We consider formalizing metadata errors to be essential for their systematic detection and correction, particularly in data integration scenarios. 

%\textbf{Formalization of temporal data errors.} We could argue that this has been already done by Josko et al. -> maybe omit it.

\textbf{Deeper investigation of data errors for different data modalities.}
While this survey focuses on tabular datasets, data errors can manifest in various forms across other data modalities, such as text, images, time series, and graphs. Future work should extend the formal definition of data errors to these modalities and investigate which error characteristics are universal versus modality-specific.

%% Possible future works! So: where can we go from here? 
% Refine refine metadata errors, temporal data errors, move to different modalities (error type definitions for other data models like NoSQL or textual data)

\begin{acks}
The authors would also like to thank Helena Galhardas and Philipp Hildebrandt for their valuable comments on this work. 
\end{acks}

\bibliographystyle{ACM-Reference-Format}
\bibliography{references}

\end{document}